\documentclass[final,3p,10pt]{elsarticle}

\usepackage{amssymb}
\usepackage{amsmath}
\usepackage{natbib}
\usepackage{lineno}
\usepackage{xcolor}
\usepackage{tabularray}
\usepackage{rotating}

\usepackage{cite}
\usepackage[colorlinks=true,linkcolor=blue,allcolors=blue]{hyperref}%

\usepackage{epsfig,color,amsmath}
 
\usepackage{wrapfig}
\usepackage{setspace}  
\usepackage{xcolor}
\usepackage{epstopdf}
\usepackage{amssymb}
\usepackage{url}
\usepackage{mathtools}
\usepackage{enumitem}
\usepackage{multirow}
\usepackage{makecell}
\usepackage{amsmath}
\usepackage{stackrel}
\usepackage{booktabs}
\usepackage[flushleft]{threeparttable}
\usepackage{makecell}

\usepackage{subfig}
\usepackage{graphicx}
\usepackage[normalem]{ulem}
\useunder{\uline}{\ul}{}
\usepackage{lscape}

\usepackage[linesnumbered,lined,boxed,commentsnumbered,ruled,longend]{algorithm2e}


\setlist[itemize]{leftmargin=*}

\makeatother

\newcommand{\mat}[1]{\boldsymbol{#1}}

\providecommand{\mF}{\ensuremath{\mat{F}}}

\newcommand{\cs}{{\cal s}}

\newcommand{\m}{\boldsymbol}
\allowdisplaybreaks[4]
\newcommand{\mr}[1]{\mathrm{#1}}

\usepackage{amsthm}

\usepackage[usestackEOL]{stackengine}
\stackMath
\usepackage[framemethod=TikZ]{mdframed}
\usepackage{bm}
\mdfdefinestyle{MyFrame}{%
	linecolor=black,
	outerlinewidth=1.5pt,
	roundcorner=1.5pt,
	innerrightmargin=5pt,
	innerleftmargin=5pt,}

\usepackage{tikz}
\usetikzlibrary{matrix,positioning,decorations.pathreplacing}

\usepackage[export]{adjustbox}
\usepackage{lipsum}
\usepackage{mathtools}
\usepackage{cuted}
\usepackage{physics}
 \usepackage{colortbl}
 \usepackage[normalem]{ulem}
\useunder{\uline}{\ul}{}
\newcolumntype{L}{>{\arraybackslash}m{3in}}

\definecolor{ceruleanblue}{rgb}{0.16, 0.32, 0.75}

\usepackage{mdframed}

\mdfdefinestyle{theoremstyle}{%
	linecolor=blue,linewidth=1pt,%
	 frametitlerule=true,%
	 frametitlebackgroundcolor=gray!20, innertopmargin=\topskip,}

  \usepackage[framemethod=TikZ]{mdframed}
\mdfsetup{skipabove=1mm,skipbelow=1mm}
\newcounter{theo}[section]
\newenvironment{theo}[1][]{%
	\stepcounter{theo}%
	\ifstrempty{#1}%
	{\mdfsetup{%
			frametitle={%
				\tikz[baseline=(current bounding box.east),outer sep=0pt]
				\node[anchor=east,rectangle,fill=cyan!30]
				{\strut };}}
	}%
	{\mdfsetup{%
			frametitle={%
				\tikz[baseline=(current bounding box.east),outer sep=0pt]
				\node[anchor=east,rectangle,fill=white!50]
				{\strut \textcolor{ceruleanblue!90}{#1}};}}%
	}%
	\mdfsetup{innertopmargin=10pt,linecolor=ceruleanblue!90,%
		linewidth=2pt,topline=true,
		frametitleaboveskip=\dimexpr-\ht\strutbox\relax,}
	\begin{mdframed}[]\relax%
	}{\end{mdframed}}

\journal{Journal of Water Process Engineering}

\begin{document}

\begin{frontmatter}



\title{Modeling and Design Optimization of Looped Water Distribution Networks using MS Excel: Developing the Open-Source X-WHAT Model}


\affiliation[1]{organization={University of São Paulo, Department of Hydraulic Engineering and Sanitation, São Carlos School of Engineering},
            addressline={Av. Trab. São Carlense, 400 - Centro}, 
            city={São Carlos},
            postcode={13566-590}, 
            state={São Paulo},
            country={Brazil}}
            
\affiliation[2]{organization={The University of Texas at San Antonio, College of Engineering and Integrated Design, School of Civil \& Environmental Engineering and Construction Management},
            addressline={One UTSA Circle, BSE 1.310}, 
            city={San Antonio},
            postcode={78249}, 
            state={Texas},
            country={United States of America}}     

\affiliation[3]{organization={Vanderbilt University, Department of Civil and Environmental Engineering},
            addressline={24th Avenue South}, 
            city={Nashville},
            postcode={37235}, 
            state={Tennessee},
            country={United States of America}}

\author[1,2]{Marcus Nóbrega {Gomes Jr.}
        }
\author[1]{Igor Matheus Benites
        }
\author[3]{Salma M. Elsherif
        }
\author[3]{Ahmad F. Taha
        }
\author[2]{Marcio H. Giacomoni
        }


\begin{abstract}
Cost-effective water distribution network design with acceptable pressure performance is crucial for the management of drinking water in cities. This paper presents a Microsoft Excel tool to model, simulate, and optimize water distribution networks with looped pipelines under steady-state incompressible flow simulations. Typically, the hardy-cross method is applied using spreadsheet calculations to estimate discharges. This method requires mass-conservative initial estimates and requires successive iterations to converge. In this paper, however, we develop an alternative method that uses the built-in solver capabilities of Excel, does not require initial mass-conservative estimation, and is free of flow corrections. The main objective of this paper is to develop an open-source accessible tool for simulating hydraulic networks also adapted for teaching and learning purposes. The governing equations and the mathematical basis for the hydraulic modeling of the system are mathematically described, considering the topology of the network, mass and energy conservation, cost of tank material, foundation, and cost of pumping energy to fill the tank. The use of this tool is encouraged at the undergraduate and graduate engineering levels, as it offers the opportunity to address complex concepts such as hydraulic modeling and network optimization in a comprehensive way using a spreadsheet that does not require coding expertise. Hence, users can debug all cells and understand all equations used in the hydraulic model, as well as modify them. To demonstrate the model capabilities, three practical examples are presented, with the first one solved step by step, and the results are compared with the EPANET and with the results reported in the literature. Using the optimization method presented in this paper, it was possible to achieve a cost reduction of 151,790 USD (9.8\% of the total cost) in a network that supplies a 44,416 population. A detailed step-by-step guide on how to use the developed tool is presented in the supplemental material.
\end{abstract}

\begin{highlights}
\item We develop a hydraulic solver for water distribution networks using Excel
\item The model does not require coding expertise and only uses built-in functions 
\item We apply the model to 3 different networks and compare results with EPANET and with reported results in the literature
\item The optimization framework can identify optimal tank water depths, potentially resulting in cost reductions of approximately 10\% compared to standard design procedure.
\item The cost function used to optimize the design considers pipeline, pump, tank material, and tank foundation costs.
\end{highlights}


\end{frontmatter}



\onehalfspacing

\section{Introduction}

Water distribution networks (WDNs) are essential infrastructure systems responsible for delivering water to residences, commercial establishments, and various other users. The efficient planning and effective design, operation, and maintenance of WDNs are critical to meet the growing demands of growing populations. Engineers and utility managers face the complex task of providing sustainable water resource management in WDNs that requires multiple factors, such as hydraulic performance and implementation and operation costs \citep{mays_water_2000,walski_advanced_2003}.

Headloss/gain models are nonlinear, which complicates decision making and network optimization. To address this, the use of specific software and tools to aid in the design of WDNs is fundamental, especially with the advance of computational resources and optimization capabilities. Such modeling packages usually perform a variety of tasks related to the design (e.g., choosing optimal network diameters), modeling, sensitivity analysis, scenario simulation, and optimization of either operation and/or implementation of WDNs \citep{awe_review_2019,sonaje_review_2015}. A wide variety of software, open source codes, and even Excel spreadsheets are available in the literature and are comprehensively described in Tab.~\ref{tab:WDN_tools}. Most of the software options listed (7 out of 11) require a paid license, which can present accessibility challenges. Although these tools offer advanced functionalities and are capable of handling larger and more complex networks, they tend to be expensive and require extensive training. The research conducted in \citep{parker_selecting_2010} states that the cost of selecting feasible and easy-to-use software is crucially important, especially in educational and research environments where multiple users can take advantage of the software package. Therefore, for smaller organizations, municipalities, or educational institutions, free or low-cost alternatives may provide feasible and sometimes more flexible solutions \citep{ku_using_2011}.

\begin{landscape}
\begin{table}[]
\caption{Water Distribution Network modeling tools and software available}
\label{tab:WDN_tools}
\resizebox{\columnwidth}{!}{%
\begin{tabular}{@{}ccllccc@{}}
\toprule
\textbf{Sotware} &
  \textbf{Cost} &
  \textbf{Main features} &
  \multicolumn{1}{c}{\textbf{Optimization   features}} &
  \textbf{Open Source} &
  \textbf{Base Platform} &
  \textbf{Development   Purpose} \\ \midrule
\multirow{5}{*}{\begin{tabular}[c]{@{}c@{}}WaterGEMS and \\ WaterCAD*\end{tabular}} &
  \multirow{5}{*}{Paid} &
  Platform interoperability, model Building, and data connection &
  Model calibration &
  \multirow{5}{*}{No} &
  \multirow{5}{*}{Standalone} &
  \multirow{5}{*}{Commercial} \\
 &
   &
  Hydraulic, operations and water quality simulations &
  Identify leak locations &
   &
   &
   \\
 &
   &
  Great data presentation and visualization &
  Optimized design and rehabilitation &
   &
   &
   \\
 &
   &
  Optimization (using genetic algorithm) &
  Optimized pump scheduling &
   &
   &
   \\
 &
   &
  Energy and capital-cost analysis &
   &
   &
   &
   \\ \midrule
\multirow{4}{*}{Pipe Flow Expert} &
  \multirow{4}{*}{Paid} &
  Intuitive user interface &
  \multirow{4}{*}{-} &
  \multirow{4}{*}{No} &
  \multirow{4}{*}{Standalone} &
  \multirow{4}{*}{Commercial} \\
 &
   &
  Piping design and pipeline modeling system &
   &
   &
   &
   \\
 &
   &
  2D and 3D isometric presentations &
   &
   &
   &
   \\
 &
   &
  Robust calculation engine &
   &
   &
   &
   \\ \midrule
\multirow{5}{*}{\begin{tabular}[c]{@{}c@{}}KYPipe\\ Pipe 2022\end{tabular}} &
  \multirow{5}{*}{Paid} &
  CAD, GIS, Google Earth Import/Export/Background Images &
  Optimal Pump Scheduling &
  \multirow{5}{*}{No} &
  \multirow{5}{*}{Standalone} &
  \multirow{5}{*}{Commercial} \\
 &
   &
  Powerful software package &
  Optimal Design Module (cost) &
   &
   &
   \\
 &
   &
  Optimization features &
  Optimized Calibration (based on filed test data) &
   &
   &
   \\
 &
   &
  Special Analyses and Network Types &
   &
   &
   &
   \\
 &
   &
  Skeletonization &
   &
   &
   &
   \\ \midrule
\multirow{5}{*}{EPANET} &
  \multirow{5}{*}{Free} &
  Powerful and versatile application (Extensively tested and reliable) &
  Only   with third party plugins &
  \multirow{5}{*}{Yes} &
  \multirow{5}{*}{Standalone} &
  \multirow{5}{*}{Educational/Research} \\
 &
   &
  Hydraulic and water quality simulation &
   &
   &
   &
   \\
 &
   &
  Integration with different platforms &
   &
   &
   &
   \\
 &
   &
  Extended period simulation &
   &
   &
   &
   \\
 &
   &
  Open source and powerful simulation and modelling engine &
   &
   &
   &
   \\ \midrule
\multirow{5}{*}{\begin{tabular}[c]{@{}c@{}}Urbano 11\\ Hydra\end{tabular}} &
  \multirow{5}{*}{Paid} &
  Google Maps import and Digital Terrain Model features &
  Pipe diameter (through genetic algorithms) &
  \multirow{5}{*}{No} &
  \multirow{5}{*}{Autodesk AutoCAD®} &
  \multirow{5}{*}{Commercial} \\
 &
   &
  Drawing and editing CAD functions &
   &
   &
   &
   \\
 &
   &
  Intersection analysis of pipe systems &
   &
   &
   &
   \\
 &
   &
  Uses EPANET engine/libraries for hydraulic calculations &
   &
   &
   &
   \\
 &
   &
  Trenches, cross sections, and excavation calculation &
   &
   &
   &
   \\ \midrule
\multirow{5}{*}{WATSYS} &
  \multirow{5}{*}{Paid} &
  Analyzing existing water supply systems &
  \multirow{5}{*}{-} &
  \multirow{5}{*}{No} &
  \multirow{5}{*}{\begin{tabular}[c]{@{}c@{}}Standalone \\ or \\ AutoCAD® integration\end{tabular}} &
  \multirow{5}{*}{Commercial} \\
 &
   &
  Deficiency Identification and Correction &
   &
   &
   &
   \\
 &
   &
  Uses EPANET engine/libraries for hydraulic and water quality analysis &
   &
   &
   &
   \\
 &
   &
  Infrastructure Maintenance Management &
   &
   &
   &
   \\
 &
   &
  Graphic Information System (GIS) Integration &
   &
   &
   &
   \\ \midrule
\multirow{5}{*}{InfoWater PRO} &
  \multirow{5}{*}{Paid} &
  Pressure zone manager &
  Many options including energy use &
  \multirow{5}{*}{No} &
  \multirow{5}{*}{\begin{tabular}[c]{@{}c@{}}Esri ArcGIS \\ or \\ AutoCAD®\end{tabular}} &
  \multirow{5}{*}{Commercial} \\
 &
   &
  Skeletonizer &
  \multirow{3}{*}{\begin{tabular}[c]{@{}l@{}}(through genetic algorithms \\ and \\ particle swarm optimization)\end{tabular}} &
   &
   &
   \\
 &
   &
  Calculations and calibrations of pipes, pumps, valves, and water quality &
   &
   &
   &
   \\
 &
   &
  Transient analysis &
   &
   &
   &
   \\
 &
   &
  What-if scenarios &
   &
   &
   &
   \\ \midrule
\multirow{5}{*}{Wadiso} &
  \multirow{5}{*}{Paid} &
  GIS Integration &
  Minimize capital expenditure &
  \multirow{5}{*}{No} &
  \multirow{5}{*}{Standalone} &
  \multirow{5}{*}{Commercial} \\
 &
   &
  Simplified model building &
   &
   &
   &
   \\
 &
   &
  Uses EPANET engine/libraries for for hydraulic calculations &
   &
   &
   &
   \\
 &
   &
  Extended Period Time Simulation and Water Quality Modelling &
   &
   &
   &
   \\
 &
   &
  Optimization Module &
   &
   &
   &
   \\ \midrule
\multirow{5}{*}{UFC System} &
  \multirow{5}{*}{\begin{tabular}[c]{@{}c@{}}Free for universities \\ and \\ paid for companies\end{tabular}} &
  Complete package for designing WDNs &
  \multirow{2}{*}{\begin{tabular}[c]{@{}l@{}}Determine diameters to obtain the most \\ cost-effective network\end{tabular}} &
  \multirow{5}{*}{No} &
  \multirow{5}{*}{\begin{tabular}[c]{@{}c@{}}AutoCAD® \\ and \\ EPANET\end{tabular}} &
  \multirow{5}{*}{\begin{tabular}[c]{@{}c@{}}Commercial\\  and \\ Educational\end{tabular}} \\
 &
   &
  Uses EPANET engine/libraries for hydraulic and water quality analysis &
   &
   &
   &
   \\
 &
   &
  Compatibilisation of water, sewage, and drainage network projects &
   &
   &
   &
   \\
 &
   &
  Generation of constructive drawings for tanks, trenches, and cross sections &
   &
   &
   &
   \\
 &
   &
  Highly detailed quantification of materials and connections &
   &
   &
   &
   \\ \midrule
\multirow{5}{*}{WDNetXL System} &
  \multirow{5}{*}{Free} &
  Advanced and robust WDN hydraulic simulation distributed as MS-Excel® add-ins &
  Multi-objective genetic algorithm &
  \multirow{5}{*}{Yes} &
  \multirow{5}{*}{Microsof Excel®} &
  \multirow{5}{*}{Educational/Research} \\
 &
   &
  Topological analysis &
  Sizing of pipes and pumping system &
   &
   &
   \\
 &
   &
  Skeletonization &
  Allocation of pressure  measurement devices &
   &
   &
   \\
 &
   &
  \multirow{2}{*}{\begin{tabular}[c]{@{}l@{}}Demand and pressure-driven analysis; Private local water storages; \\ Multi-floor buildings; Remote control of pressure reduction\end{tabular}} &
  Design of segments/modules &
   &
   &
   \\
 &
   &
   &
   &
   &
   &
   \\ \midrule
\multirow{4}{*}{EPANET-Matlab-Toolkit} &
  \multirow{4}{*}{Free} &
  MATLAB environment that provides a programming interface with EPANET &
  \multirow{4}{*}{\begin{tabular}[c]{@{}l@{}}There are no built-in functions, but a wide range of   \\ optimization tools available in the MATLAB \\ environment can be used\end{tabular}} &
  \multirow{4}{*}{Yes} &
  \multirow{4}{*}{MATLAB®} &
  \multirow{4}{*}{Educational/Research} \\
 &
   &
  View, modify, simulate, and plot EPANET library results &
   &
   &
   &
   \\
 &
   &
  Modular architecture allows to use of the functions in other Matlab   projects &
   &
   &
   &
   \\
 &
   &
  Uses EPANET engine/libraries for hydraulic and water quality analysis &
   &
   &
   &
   \\ \midrule
\multicolumn{7}{l}{* WaterCAD is a   subset of WaterGEMS.} \\ \bottomrule
\end{tabular}%
}
\end{table}
\end{landscape}

Research conducted in \citep{thakur_limited_2012} and \citep{von_krogh_open_2007} argue that open-source software plays an important role in facilitating collaboration, promoting innovation, and providing accessible tools, which has implications for encouraging innovation in any field or industry. For instance, the Environmental Protection Agency Network Evaluation Tool (EPANET) \citep{rossman_epanet_2000} has been a critical asset in the modeling of WDNs for more than two decades. Its importance lies in its accessibility and powerful capabilities. As an open-source tool, it is freely available, making it a valuable resource not only for regular users but also for advanced users who can use the open-source availability to adapt to more complex scenarios and simulations. Several other free and paid software tools use the EPANET Engine/Libraries Toolkit to perform hydraulic and water quality calculations (see Table~\ref{tab:WDN_tools}). 

In terms of development platform, commercial software options consist of either a proprietary standalone platform or integration with Autodesk AutoCAD. Among the three open source tools listed in Tab.~\ref{tab:WDN_tools}, one uses the Microsoft Excel environment \citep{giustolisi_excel-based_2011} while another uses the MATLAB environment \citep{eliades_EPANET-matlab_2016}. The study developed in \citep{eliades_EPANET-matlab_2016} presents an open-source software connecting MATLAB to EPANET, in an intuitive and relatively easy-to-use way. The Toolkit allows the user to access EPANET through their shared object libraries, as well as their executables. In summary, their application provides a programming interface with EPANET directly from the MATLAB environment, where it is possible to visualize, modify, and plot results from the EPANET library. Furthermore, its modular architecture makes it possible to use the functions in other MATLAB projects.

Another example is the WDNetXL that is developed in \citep{giustolisi_excel-based_2011}. It is an advanced tool that uses MS-Excel add-ins for WDNs analysis including hydraulic simulation, valve system analysis, pipe failure scenarios, optimal pipe sizing, and optimal valve system design. Although listed as free in Tab.~\ref{tab:WDN_tools}, the software is no longer available in their website to the date of this paper. Several studies have shown that Excel-based tools can be effective in conveying complex hydraulic concepts and principles in an easy-to-understand comprehensive way. For instance, the study conducted in \citep{brkic_spreadsheet-based_2017} has shown that Excel is useful in analyzing WDNs and can help students understand the fundamental concepts and principles of hydraulic network analysis and design. Similarly, another study presented in \citep{huddleston_water_2004} proposes a method of using Excel for WDNs analysis based on linear theory, highlighting its accessibility and efficiency in teaching and learning. Furthermore, authors in \citep{rivas_application_2006} emphasize Excel's role in introducing students to concepts in thermodynamics and hydraulic engineering. Their results show the performance of Excel as an alternative tool to simulate relatively complex problems, not only for solving real-world problems but also for educational purposes. 

Other studies have addressed similar issues. However, they have focused on the use of Excel solver for analysis of WDNs. Research conducted in \citep{adedeji_spreadsheet_2017} demonstrates the effectiveness of using the Hardy-Cross \citep{cross1936analysis} method with the solver to estimate pipe flow. They present an example of a network with a single loop, demonstrating how efficiently the solver optimizes flow correction factors for a relatively small WDN. In addition, \citep{jewell_teaching_2001} uses the solver and a studio classroom setting to teach hydraulic design for water distribution systems and open channel flow and has concluded that these tools can improve student learning and engagement. Research presented in \citep{oke_statistical_2015} recommends Microsoft Excel Solver as an optimal approach to solving linear equations in pipe network analysis. Furthermore, \citep{gokyay_easy_2020} presents a Visual Basic Application (VBA) that implements the Hardy-Cross algorithm to solve a given WDN by balancing head losses in a looped network system. In addition, the study \citep{demir_ms_2018} shows an MS Excel tool, called YTUSU, developed for teaching hydraulic analysis and aiding the design of water distribution networks in environmental engineering education. However, these studies exclusively use the Hardy Cross method within an Excel environment. In other words, they manually implement it through a trial-and-error approach rather than providing a dedicated tool or software solution.

It is noteworthy that several studies \citep{barron_corvera_herramienta_2021,niazkar_analysis_2017,niazkar_analysis_2017-1,turkkan_visual_2020,wahba_improved_2015} have explored performing these calculations using tools beyond Excel, including programming languages like MATLAB and C\#. Excel is a powerful software that is widely used in different applications. It serves a multitude of functions, including basic calculations, data analysis, design, and visualization. Using Excel to model and optimize WDNs has several advantages. First, it is a cost-effective tool. Second, its user-friendly interface ensures accessibility to individuals, including students and engineers, with limited programming experience. Third, its license cost is relatively low compared to dedicated commercial WDN design software and users can utilize Excel for a variety of other applications rather than just hydraulic simulation, making the license cost virtually negligible. 

\subsection{Paper objectives and Contributions}
As seen above, several studies have explored the use of Excel and MATLAB to model and analyze WDNs. However, most of these studies focus only on the implementation of the Hardy-Cross algorithm in these environments that require successive iterations and initial mass-conservative flow estimations in all network nodes. In addition, currently, there is no publicly available tool in Excel that offers a rapid and easy-to-access tool for hydraulic simulation and preliminary design of WDNs, whether for applying it in classroom examples or in real-world case studies. Therefore, this paper introduces a new Microsoft Excel tool that can be used to design, simulate, and optimize tank design used to provide water to water distribution networks under a variety of network topology and boundary conditions. 


The fundamental contributions of this paper are:
\begin{itemize}
    \item We developed an open-source model, the Excel for Water system Hydraulic Analysis Tool  (X-WHAT) to aid in the simulation and design of WDNs.
    \item We develop a novel WDN cost function that considers more realistic factors rather than just the hydraulic features of the WDN, such as pipeline costs, pumping costs associated with filling the tanks, and tank material and foundation costs. 
    \item The foundation cost function used in this paper considers structural design factors, such as lateral non-linear wind forces, for ground and elevated tanks.
    \item The methodology presented in the numerical model herein developed merges fundamental concepts of hydraulics and structural design, allowing an interdisciplinary application for teaching in engineering classes.
    \item The methods presented in this paper are solved in a simple Excel spreadsheet, without requiring any other software package, and do not require coding expertise since the model uses only built-in Excel functions. 
\end{itemize}


The remainder of the paper is structured into five sections. Section \ref{sec:mat_methods} presents the fundamental equations used in the model, including the conservation of mass in junctions, tanks, tanks, and pumps. In addition, it details the methods to compute the conservation of energy in loops. The estimation of the forces acting on the foundation and the function describing the foundation cost is also explained in this section. Finally, two optimization problems are formulated to deal with the problem of finding flows in a given WDN and optimizing the tanks design given a set of constraints. Section \ref{sec:case_studies} presents three numerical case studies where the methods and the developed tool are applied. Section \ref{sec:results} and Section \ref{sec:discussion} present the results and discussion, respectively. Finally, Section \ref{sec:conclusions} presents the conclusions of the paper.

\section{Materials and Methods} \label{sec:mat_methods}
In this section, we provide a brief introduction and explanation of the governing equations, conservation laws, methods, and materials used to build our models and obtain the subsequent results. 

\subsection{WDNs Components and Topology}
The main elements of the pressurized looped hydraulic networks under steady-state flow distribution of an incompressible fluid are the nodes and links. Nodes include tanks, junctions, and tanks, while links include pipes, pumps, and valves. They can have known boundary conditions, such as observed heads of and flows. A junction connects two or more links and is defined by an elevation value from a reference datum. For the steady-state simulation, tanks and tanks are assumed to have a fixed piezometric pressure (known or optimized in the algorithm), such that their levels are assumed to be constant. Therefore, for the sake of the WDN hydraulic simulation, tanks or tanks are defined by their piezometric head at their surface. Tanks are storage elements with changing volumes and heads but since we are only modeling the steady-state case for the so-called design conditions, tanks can also be assumed with a fixed head that could represent a critical condition for the tank. 

In addition, a pipe is a conveying element having a constant slope and is defined by the pipe diameter ($D$), pipe length ($L$), and friction properties that are later defined in the following sections. The pump elements, as well as valves, can be used to control the network dynamics \citep{creaco2019real}. However, in this paper, pumps are only considered to be connected to tanks or tanks to fill them up to the required level. Therefore, for a given tank or tank known head, the pumps do not alter the network dynamics; nonetheless, the pump energy costs are associated with the tank heads, which ultimately are directly connected to the network performance to maintain acceptable heads, velocities, and pressures. The smaller the designed tank head, the lower the pump energy costs; however, the higher the chances of having a storage node that provides the minimum required pressure in the network. 

\begin{figure}
    \centering
    \includegraphics[scale = 0.12]{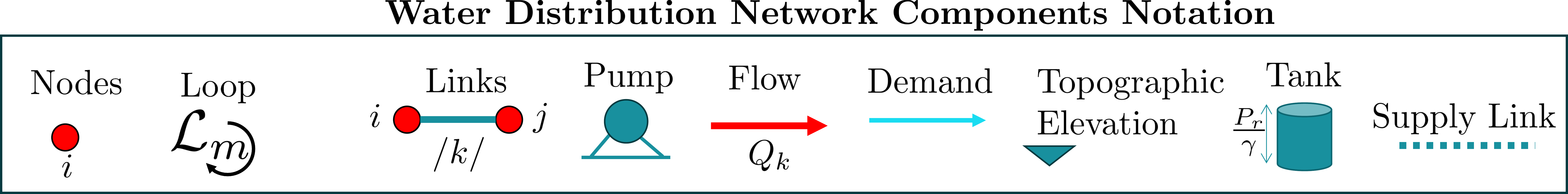}
    \caption{Notation used in the paper for describing the WDNs Components. Nodes and links are continuous numbered. Loops follow the clockwise convection. Nodes, tanks, and pumps are defined by their topographic elevation. Tanks have their dynamic pressure defined by $P_{\mr{r}}/\gamma$.}
    \label{fig:notation}
\end{figure}

In this paper, and in the models developed herein, we focus only on junctions, links, tanks, and pumps. We assume that the valves are fully open and we neglect the minor losses in the network. Moreover, we do not attempt to use pumps as a controlling factor of network dynamics; rather, we consider the pumps to calculate the associated energy costs that influence the overall cost design optimization of the whole network. The notation used in the paper to describe the network components is shown in Fig.~\ref{fig:notation}.

\subsection{Conservation Laws}
WDNs are governed by laws of conservation of mass and energy, described in this section.

\subsubsection{Conservation of Mass}
The principle of mass conservation ensures that the continuity equation is satisfied for all nodes; the change in storage is null (i.e., steady-state) and is equal to the algebraic difference between the inflows and outflows. 


\paragraph{Junctions} They are considered elements with no storage volume connecting one or more links and allow direct demand withdraws from them. 
The conservation of mass for junctions is expressed as:

\begin{equation}~\label{equ:JuncMass}
     \sum_{k = 1}^{n_k}  Q_{\mr{in}}^{i} - \sum_{m = 1}^{n_m} Q_{\mr{o}}^{i} - Q_\mathrm{d}^i = 0,~\text{For junctions from i = 1, 2, $\cdots$ $n_d$}
\end{equation}
where $Q_{\mr{i}}$ are the inflows and $Q_{\mr{o}}$ the outflows. The counters $n_k$ and $n_m$ represent the number of pipes entering and leaving the node $i$. The demands in each junction are given by $q_\mathrm{d}$ where positive values represent water consumption in the network. 

\paragraph{Reservoirs} They are assumed to be an infinite source of water with fixed heads, $h_\mathrm{r} = \mathrm{constant}$. However, in a design optimization simulation, $h_{\mr{r}}$ can be estimated to minimize network costs.

\paragraph{Tanks} Although tanks can have a varying water depth under unsteady state simulations, since we are only modeling the steady-state dynamics, we consider a fixed head in tanks. From now on, tanks and reservoirs are considered interchangeably.

\paragraph{Pumping Flow Rate}
Although all the equations developed in this paper are defined for a steady-state case where demands and supply are equal, the pumping flow rates are different from the supply released by the tank in the WDN simulation. The hour factor in the demand ($k_2$), as well as the duration at which the pumps work ($n_p$) are considered to calculate the pump flow rate ($Q_{\mr{p,r}}$), which can be calculated as:

\begin{equation} \label{equ:pump_flow}
    Q^i_{\mr{p,r}} = \frac{Q^i_d}{k_2}\Bigl (\frac{n_{\mr{WDN}}}{n^i_{\mr{p}}} \Bigr),~\text{For tanks/reservoirs from $i = 1, 2, \cdots n_{\mr{r}}$}
\end{equation}
where $Q_d$ is the tank outflow rate, $n_{\mr{WDN}}$ is the duration per day ($\leq 24~\mr{h}$) that the water distribution network is operating (i.e., typically 24 h) and $k_2$ is a statistical factor related to the average demand with the maximum expected demand, herein assumed as 1.5.

\subsubsection{Conservation of Energy}
Conservation of energy over links is classified into two main principles, energy loss and energy gain. Energy loss is either due to friction loss distributed over pipes' length or localized at sources of change (e.g., valves and bends). In our paper, we neglect the effect of the localized energy losses and convert valves' losses into equivalent friction losses \citep{whiteFluidMechanics1966a}. On another hand, energy gain is provided by pumps---more details on pumps operation are briefly explained in the following sections.

\paragraph{Head Loss}
Head loss models are non-linear functions dependent of the flow discharge in a pipe $(q_o^i)$ due to the friction between the fluid and the walls of the conduit. The discharge in a pipe is now denoted by $Q$. The friction loss equation can be expressed as:

\begin{equation} \label{equ:friction_model}
    \Delta h_f = k Q | Q |^ {n-1}
\end{equation} 
where $\Delta h_f$ is the head drop due to friction losses, $k$ is a term dependent on the link properties such as diameter, length, and rugosity. The exponent $n$ is a parameter that depends on the friction model used, and $Q$ is the flow rate in the link. Minor losses in junctions are not considered. In the following sections, we introduce two friction models and the corresponding derivation of $k$ and $n$.

\textit{-- Darcy-Weisbach (D-W) Friction Model.}
The physically-based head loss model of Darcy-Weisbach is given by
\begin{equation} \label{equ:D_W}
    \Delta h_f \Big|_\text{D-W} = \frac{fLv^2}{2gD},
\end{equation}
where $f$ is the friction factor, $v$ is the average flow velocity in the pipe, $D$ and $L$ are the pipe's diameter and length, and $g$ is the gravity acceleration.

Rearranging Eq.~\eqref{equ:D_W} following Eq.~\eqref{equ:friction_model} and using $v = Q/(\pi D^ 2/ 4)$, we can derive that:
\begin{equation}~\label{equ:kDW}
    k \Big|_\text{D-W} = \frac{8 f L}{D^5 \pi^2 g},\; ~ n = 2.
\end{equation}

Several formulations for the friction factor are available in the literature. Some of them are implicit; others are limited in terms of the Reynolds number (Re), such as the Swamee-Jain formulation \citep{rossman_epanet_2000}. A more recent formulation, however, allows an explicit and not bounded by Reynolds number and can be written as \citep{porto2004hidraulica}:


\begin{equation} \label{equ:friction_factor}
    f=\Bigg(\left(\frac{64}{\mathrm{Re}}\right)^8+9.5\left[\operatorname{ln}\left(\frac{\epsilon}{3.7 D}+\frac{5,74}{\mathrm{Re}^{0.9}}\right)-\left(\frac{2500}{\mathrm{Re}}\right)\right]^{-16}\Bigg)^{0.125}.
\end{equation}
where $\epsilon$ is the length of the rugosity and $D$ is the diameter. $Re$ is the Reynolds number, defined by $Re = \frac{v D}{\nu}$, where $v$ is the flow velocity and $\nu$ the kinematic viscosity. 


\textit{-- Hazzen-Willians (H-W) Friction Model.}
The Hazzen-William equation is considered an empirical model, where $k$ and $n$ in Eq.~\eqref{equ:friction_model} are expressed as:

\begin{equation}~\label{equ:kHW}
    k \Big|_\text{H-W} = \frac{10.67}{C^{1.85}D^{4.87}},\; ~ n = 1.85,
\end{equation}
where $C$ is the H-W friction coefficient and depends on the material of the pipe. 

\paragraph{Node Pressures}
Within 2 consecutive nodes (i.e., a link), we can apply the Bernoulli's equation to determine the heads in each node such that:

\begin{equation}
    h_{\mr i} - h_{\mr j}  =  k_{{f}}^{\mr l} |Q_{\mr l}| Q_{\mr l}^{n-1},~\text{For all links from $i$ = 1,~2,$\cdots$,~$n_l$}
\end{equation}
where $h_{\mr i}$ and $h_{\mr j}$ := ($P/\gamma + z)$ are the piezometric heads at nodes $i$ and $j$ 

The pressures heads of the nodes can hence be calculated by subtracting the elevation level ($z$) from the head.

\begin{figure}
    \centering
    \includegraphics[scale = 0.2]{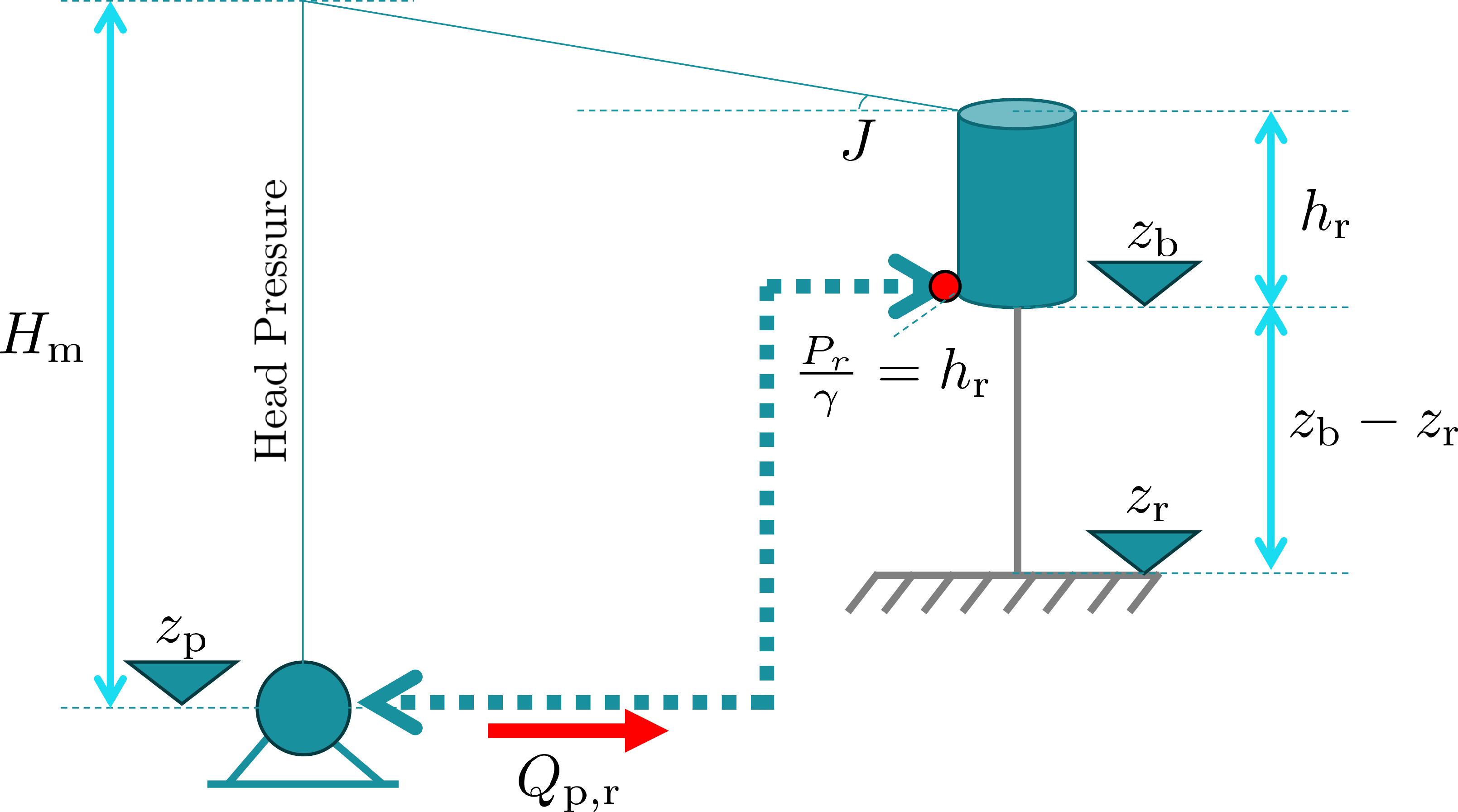}
    \caption{Pump to tank scheme, where the manometric head gain $H_{\mr{m}}$ is calculated accounting for the head losses in the pipe and the geometric differences within the pump level ($z_{\mr{p}}$) and the bottom of the tank $z_{\mr{b}}$.}
    \label{fig:pump_reservoir}
\end{figure}

\paragraph{Head Gain}
Each WDN has its unique topology and layout. Although pumps can actively change the WDN dynamics by adding pressure head to the nodes, in this paper, we only consider pumps that provide water to tanks. The network is driven by the gravitational gradients from the tank that are filled by pumps. Therefore, pumping in our model, does not change the pressure and flow dynamics in the network; however, they influence the overall cost of the network, especially if relatively high manometric pressures are required to pump the water into the tanks. A schematic of the pump system to provide water to tanks is shown in Fig.~\ref{fig:pump_reservoir}

The pumps have a manometric pressure $H_{m}^p$ and a friction coefficient $k^p$. The manometric head gain $(H_{\mr{p}})$ can be calculated considering the pump elevation $(z_{\mr{p}})$, the geometric elevation within the pump and the tank ($h_{\mr{g}}$), and accounting for the linear friction loss in the pipe connecting the pump and the tank, such that:




\begin{equation}
    H_{\mr{p}} = h_{\mr{g}} + k^p Q_{\mr{p,r}} \abs{Q_{\mr{p,r}}}^{n-1}
\end{equation}

Note that the geometric elevation within the pump and the tank plus the pressure head at the entrance of the tank ($h_{\mr{g}}$) depends on the tank head ($h_{\mr{r}}$) and is calculated as $h_{\mr{g}} = h_{\mr{r}} + (z_b - z_{\mr{p}})$. 

\subsubsection{Loop Energy Conservation}
Let $c$ be the number of loops in the network system. Given a junction $i$, the sum of all head losses within the network returning to node $i$ must be zero to maintain energy conservation. Each loop in the network results in one conservation of energy equation. For given loop $v$ containing $m$ pipes defined by $v_1,~v_2, \cdots v_m$ , and assuming an arbitrary flow convention (e.g., clockwise), one can assign which pipes are in the expected flow direction or not by the signal of the flow discharge, resulting in the energy conservation at loop $v$ as follows:

\begin{equation} 
k_{v_1} Q_{v_1} |Q_{v_1}|^{n-1} + k_{v_2} Q_{v_2} |Q_{v_2}|^{n-1} \cdots + k_{v_m} Q_{v_m} |Q_{v_m}|^{n-1} = 0
\end{equation}

Expanding for $\mathcal{L}$ loops, the equations can be rearranged in a non-linear system of equations summarized in vector form as:

\begin{equation} \label{equ:energy_conservation}
    0=\sum_{i = 1}^{m_{v}}k_{{v_i}} Q_{v_i}\left|Q_{v_i} \right|^{n-1},~\text{For all loops from $v = 1$ to ${c}$}
\end{equation}
where $m_{v}$ is the number of pipes in loop $v$.

Eq.~\eqref{equ:energy_conservation} can be solved in state-space format by having a topologic matrix that collects the pipes (i.e., columns) in terms of each loop (i.e., rows), with a value of 1 given for flows following the clockwise convention, 0 for pipes that do not belong to the loop, and -1 for pipes that follow the convention. 

\subsection{Incidence Matrices}
The topology of the problem relates pipes and nodes by defining two representative matrices: $\m F_d$ and $\mF_l$. The matrix $\m F_d$ represents the connection of every node with respect to the links, such that:

\begin{equation}
{\mathbf{F}}_{d}(i, j)= \begin{cases}-1 & \text { if pipe } j \text { leaves juntion } i \\ 0 & \text { if pipe } j \text { is not connected to juntion } i \\ +1 & \text { if pipe } j \text { enters juntion } i\end{cases}
\end{equation}

By multiplying this matrix with the decision vector for discharges $\m q$ and subtracting by the node demands $\m d$, we define the first constraint to the problem, given in Eq.~\eqref{equ:JuncMass}.

The other matrix required to solve the fundamental conservation equations is the loop incidence matrix ($\m F_l$), given by:

\begin{equation}
\mathbf{F}_{l}(i, j)= \begin{cases}-1 & \text { if pipe } j \text { is in loop } i \text { and their directions are opposed } \\ 0 & \text { if pipe } j \text { is notin loop } i \\ +1 & \text { if pipe } j \text { is in loop } i \text { and their directions }\end{cases}
\end{equation}

Using $\m F_l$ and the flow discharges $\m q$, we solve the energy conservation in all loops given by Eq.~\eqref{equ:energy_conservation}.

\subsection{Cost Functions}
Typically, studies developing optimization techniques to enhance the design of WDN consider only the pipeline costs. In this paper, we include three new associated costs, in addition to the pipeline costs, to define the overall WDN cost: the tank material, the tank foundation, and the pumping energy costs. The functions to calculate these costs are described in this section.

\subsubsection{Tank Costs}
\paragraph{Tank Material}
The tank material cost is composed of the lateral, bottom, and top areas, and is a function of the cost of the tank material, given by:

\begin{equation} \label{equ:reservoir_material}
    C_{\mr r}^{\mr m} = c_\mr{m} \Bigl(\pi D_{\mr r}(D_{\mr r}^2/2+h_{\mr r})\Bigr)
\end{equation}
where $c_{\mr{m}}$ [$\mr{USD} \cdot \mr{L}^{-2}$] is the cost per square unit of the tank material.

\begin{figure}
    \centering
    \includegraphics[scale = 0.20]{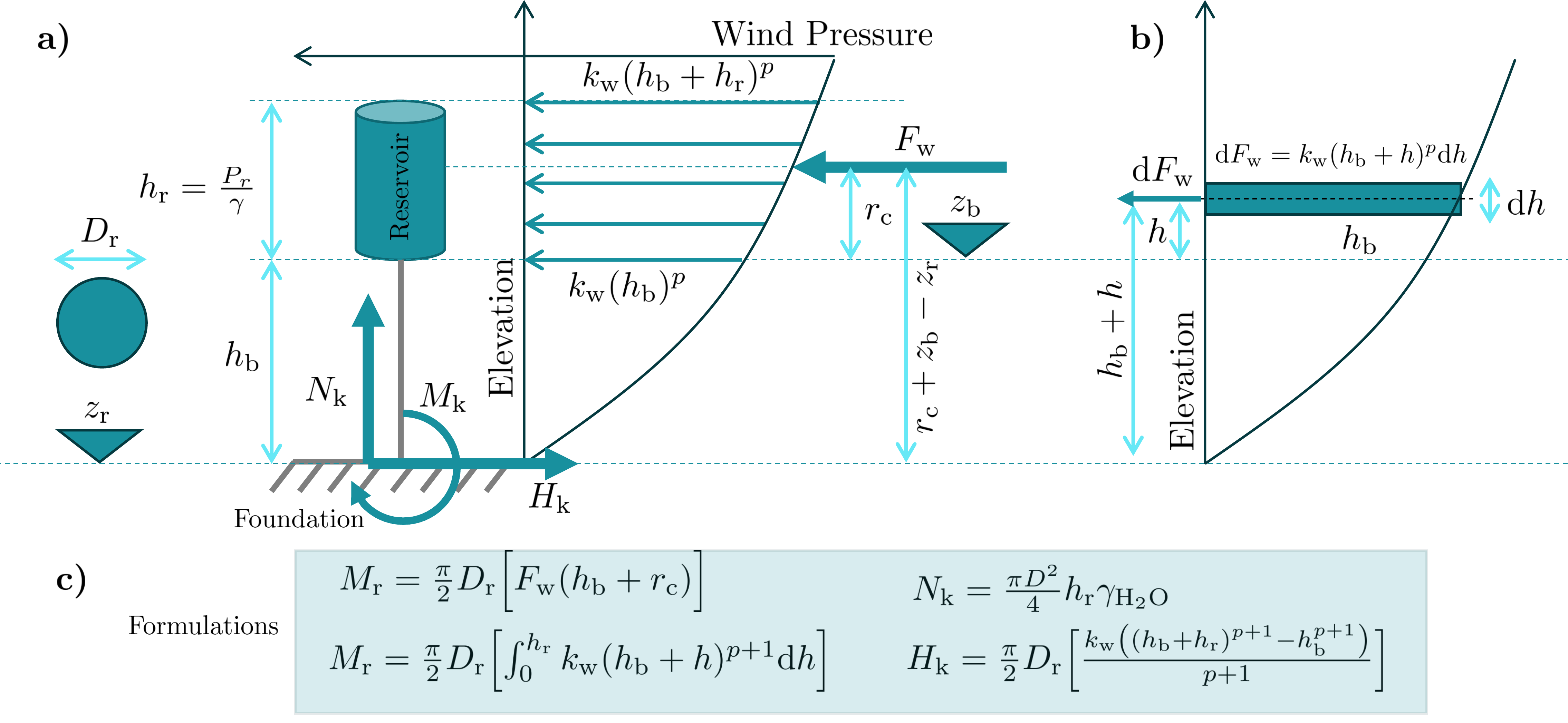}
    \caption{Static representation of the lateral forces acting in the tank surface, where (a) shows the lateral forces acting in the tank. The hydrodynamic force is modeled through $k_{\mathrm{w}} (h_b + h)^p$, with $k_{\mr w}$ being wind drag coefficient [$\mr{F}\cdot \mr{L}^{-1(2+p)}$], resulting in a force per unit elevation [$\mr{F}\cdot \mr{L}^{-1}$]. This force is integrated to produce $F_{\mr w}$ that represents the total force per unit of tank width that if multiplied by the diameter of the tank, gives the total lateral force at the foundation. This lateral force can be used with the lever arm $r_c$ to calculate the bending moment at the foundation. Alternatively, the total bending moment using static moment formulation is presented in (b), where $M_{\mr r}$ is the total bending moment projected at the tank lateral surface as a function of the tank diameter and pressures. At the foundation level, the normal force is the tank weight, and the horizontal force equals $F_{\mr k}$. The variable $r_c$ is the lever arm and $\gamma_{\mr{H_2O}}$ is the specific weight of the water.}.
    \label{fig:reservoir_static}
\end{figure}

\paragraph{Tank Foundation}
A simplified schematic representation of the forces acting on the lateral surface of the tank is presented in Fig.~\ref{fig:reservoir_static}. Other forces, rather than the lateral wind forces acting at the tank, can also be important, such as the hydrostatic pressure against the wind direction, which would reduce the foundation bending moment and the weight of the pillars, resulting in a more complex formulation for an explicit cost definition. Herein we attempt to focus on the most important one, which, for elevated tanks, is the wind forces acting at the lateral surface of the tank. The bending moment at the foundation is derived, neglecting the reducing effect of the hydrostatic pressure. The resulting general spandrel formulation is:

\begin{equation}
    M_{\mr r} = \frac{1}{2}\pi D_{\mr r} k_{\mr w}\Bigl[\frac{(h_{\mr b}+ h_{\mr r})^{p+2}-h_{\mr b}^{p+2}}{p+2} \Bigr]
\end{equation}
where $k_{\mr w}$ is the wind dynamic drag coefficient [$\mr{F\cdot L^{-(2+p)}}$] that depends on the air density, the expected wind speed and $p$ is an exponent that increases with the topographic elevation and can be estimated in terms of the wind speed $v_{\mr w}$, such that $k_{\mr w} = 0.613\frac{0.75^2 \mr{v}_{\mr w}^2}{10^p}$, with $v_{\mr k}$ in [$\mr{m}\cdot \mr s^{-1}$] and $k_{\mr w}$ in [$\mr{N}\cdot \mr{m}^{-(2+p)}$]. Values of p vary between 0.12 to 0.035 with mean values around 0.20, and they shape the curvature of the lateral forces in the tank, as shown in Fig.~\ref{fig:reservoir_static}. For a wind velocity of $\mr{40}~\mr{m\cdot s^{-1}}$ and $p = 0.30$, $k_{\mr w} = 348~\mr{N}\cdot \mr{m}^{-2.3}$

\begin{figure}
    \centering
    \includegraphics[scale = 0.3]{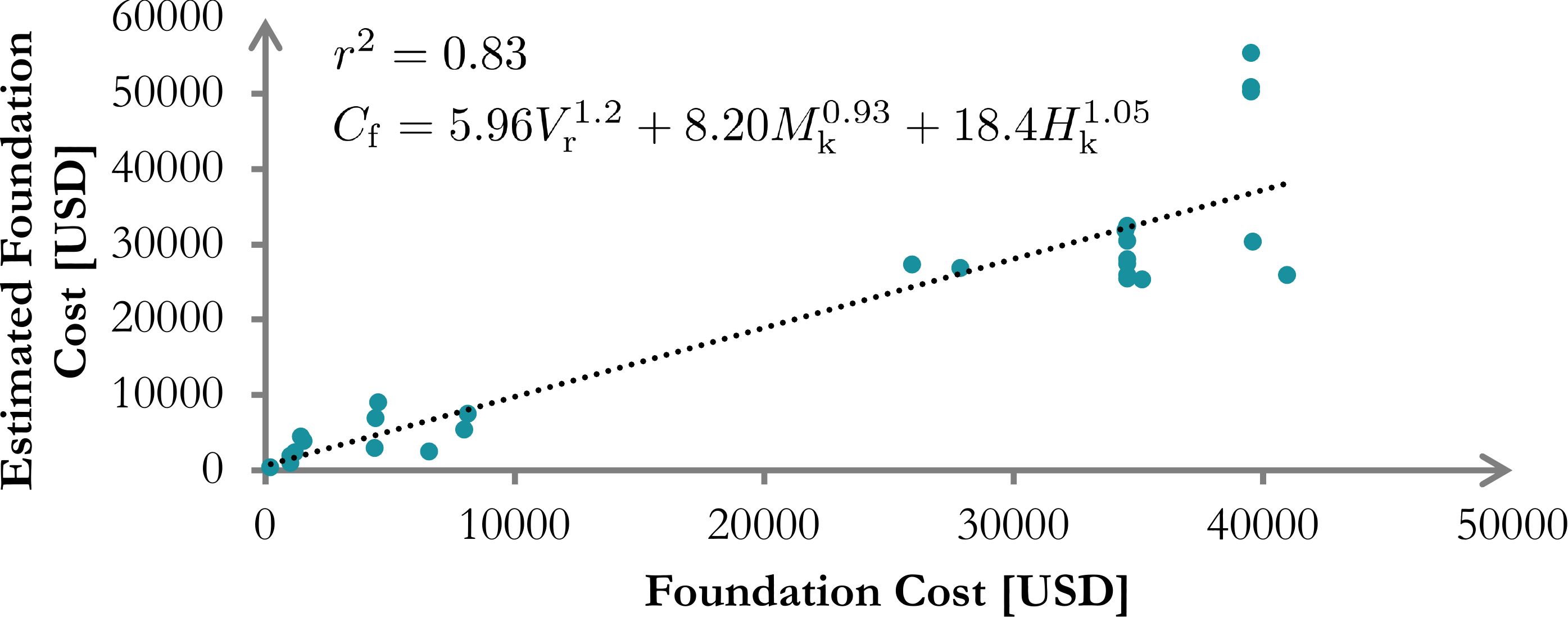}
    \caption{Foundation cost model in comparison with a regression model, where a structural foundation model was run for combinations of 10, 100, and 1000 $\mr{m^3}$, $\mr{kN.m}$, or $\mr{kN}$ of tank volume, lateral bending moment, and horizontal wind force. The regression parameters of Eq.~\eqref{equ:foundation_costs} are presented in the figure.}
    \label{fig:foundation_costs}
\end{figure}

We run a structural reinforced concrete footing model to calculate the concrete and rebar costs for different values of tank volume, height, and elevation. Effects such as tipping, sliding, and concrete shear resistance are considered, and the minimum cost design is searched. Details of the foundation model can be found in the supplemental material. We assume that the cost of the foundation is a function of the forces and bending moments at the surface of the foundation, such that:

\begin{equation} \label{equ:foundation_costs}
    C_{\mr f}^{\mr r} = \alpha_1 V_r^{\beta_1} + \alpha_2 M_k^{\beta_2} + \alpha_3 H_k^{\beta_3}
\end{equation}
where all parameters are fitted using a dataset of varied values of strengths. The results of the varied simulation tested (each strength value from 10, 100, 1000 $\mr{kN}$ or $\mr{kN.m}$) are shown in Fig.~\ref{fig:foundation_costs} and the parameters in equation \eqref{equ:foundation_costs}.

\paragraph{Tank Total Cost}

The total tank cost is the sum of the material cost (Eq.~\eqref{equ:reservoir_material}) and the foundation cost (Eq.~\eqref{equ:foundation_costs}):

\begin{equation} \label{equ:reservoir_cost}
    C_{\mr r}(V_{\mr r},z_{\mr b}, h_{\mr r}) = C_{\mr r}^{\mr m}(V_{\mr r},~h_{\mr r}) + C_{\mr{r}}^{\mr f}(V_{\mr r},h_{\mr b},h_{\mr r})
\end{equation}
where $C_{\mr r}^{\mr f}$ is the foundation cost.

In WDNs with only one tank, the total water demand is only from it. However, in networks with more than one tank, a decision is made to distribute the volumes of the tank based on zonation studies such that each tank would make available a percentage of the total volume. Alternatively, a linear relationship between the water heads or water volumes of the tanks can be used in the design \citep{huddleston_water_2004}.

To solve Eq.~\eqref{equ:reservoir_cost}, we can derive a relationship between $D_{\mr r}$ and $h_{\mr r}$ by using the tank volume and use this relationship to reduce the number of decision variables in the problem. The tank volume is typically defined by local regulations. In this paper, we assume the tank volumes as 1/3 of the total demanded volume on the day of the highest demand within the period ($V^i_{\mr r} = 1/3 k_1 \sum d^i 86400$), thus, for a given tank depth $h_{\mr r}$, the diameter is:

\begin{equation}
    D_{\mr r} = \sqrt{\frac{4 V_{\mr r}}{\pi h_{\mr r}}}
\end{equation}
where the tank volume $V_{\mr r}$ is calculated with the demanded volume of the tank within one day.

\subsubsection{Pump Energy Cost}
The pumping energy cost depends on the pump gain, the average daily duration that the pump is operated, the cost of kWh of energy, and  the pump efficiency, such that:

\begin{equation} \label{equ:daily_energy_pump}
    C_e = P_e \overbrace{\frac{H_{\mr{p}} Q_{\mathrm{p,r}} \gamma n_h} {\eta_{\mr{p}}}}^{E_{\mr{p}}}
\end{equation}
where $C_e$ is the daily cost of energy, $P_e$ is the average energy cost ($\mr{USD/kWh}$), $\eta_h$ is the pump daily number of working hours, $\eta_{\mr{p}}$ is the pump efficiency, and $E_{\mr{p}}$ is the pump daily energy consumption ($\mr{kWh}$).

The pump energy cost, as opposed to the pipeline cost and tank cost, which are capital investments, is charged throughout the years. We can define the pump energy cost in terms of the net present value. To this end, the useful life of the network ($n_l$) and the interest rate ($i_{\mr{r}}$) of the project, as well as the annual increase rate in the energy price ($i_e$) are considered to calculate the net present value (NPV) of the pumping energy, such that the conversion rate is given by:

\begin{equation}
I_{\mr{NPV}}=\left\{\begin{array}{cc}
{\left[\frac{1-\left(1+i_e\right)^{n_{\mr l}}\left(1+i_{\mr{r}}\right)^{-n_l}}{i_{\mr{r}}-i_e}\right]} & \text { if } i_{\mr{r}} \neq i_e \\
{\left[\frac{n_l}{1+i_{\mr{r}}}\right]} & \text { if } i_{\mr{r}}=i_e
\end{array}\right.
\end{equation}
where $I_{\mr{NPV}}$ [$\mr{year^{-1}}$] converts the total annual energy costs during the lifespan ($E_a$) into a present value that can be summed with the network implementation costs. One can estimate the annual energy cost by multiplying $C_e$ (i.e., the daily pumping cost from Eq.~\eqref{equ:daily_energy_pump}) by 365 days, such that the pump net present value cost ($C_{\mr{p}}$) is:

\begin{equation} \label{equ:pump_cost}
    C_{\mr{p}} = I_{\mr{NPV}} C_e \times 365
\end{equation}

\subsubsection{Pipeline Cost}
The implementation cost of pipelines is typically a function of the pipe diameters. We approximate the pipeline costs using a high-degree polynomial equation \citep{bragalli2012optimal} (i.e., herein we assume a 7-order polynomial), expressing the pipeline cost as:

\begin{equation} \label{equ:pipeline_cost}
    C_l(D) = \sum_{s = 1}^{n_d} \alpha_l D^{s}
\end{equation}
where $\alpha_l$ are the least square fit parameters that can be derived in Excel using functions such as $\mathtt{LINEST}$. 

\subsection{WDN Overall Cost}
Combining Eq.~\eqref{equ:pipeline_cost}, Eq.~\eqref{equ:pump_cost}, and Eq.~\eqref{equ:reservoir_cost} we can define the design optimization cost function given by:

\begin{equation} \label{equ:overall_cost}
    C_t = \sum_{i = 1}^{n_l} L_i C_l + \sum_{j = 1}^{n_{\mr{p}}} C_{\mr{p}}^j  + \sum_{k = 1}^{n_{\mr{r}}} C_{\mr{r}}^k  
\end{equation}
where $C_t$ is the total net present value of the network, $L_i$ is the pipeline length for link $i$, $n_l$ is the number of links, $n_{\mr{p}}$ is the number of pumps, assumed equals the number of tanks $n_{\mr{r}}$, and $h_{\mr{r}}$ is the tank head.

\subsection{Water Flow Problem (WFP)}
Two problems are formulated. The first problem (WDN Water Flow Problem) is a hydraulic simulation of a given network with known pipe diameters and tank boundary conditions and consists of finding the discharges and flow directions \citep{wang2020new}. This problem doesn't require an objective function and the solution to this problem is a vector that satisfies a system of linear and non-linear equations. This problem is relatively less complex to solve in classes, as opposed to typical approaches that use the Hardy-Cross method, because it doesn't require an initial, mass-balance conservative, estimations of the flows in all pipes, which can be time-consuming  \citep{huddleston_water_2004}. The lack of mass-conservative initial values is a major advantage of using a matrix-wise formulation approach via optimization, although mass-conservation solutions are desired to accelerate gradient-based solvers.

The second problem (WDN Design Optimization Modeling) is, for a given network with known diameters, to find not only the correct flow discharges that satisfy energy and mass conservation laws but also minimize the network's overall costs while being constrained by maximum and minimum pressure boundaries. This problem is more oriented toward a real-world engineering application as it uses hydraulic simulation as a tool to find cost-effective solutions. Following, the two problems are mathematically described.

\paragraph{WDP Water Flow Problem}
This problem is mathematically written as:

\begin{theo}[\textbf{Water Flow Problem}]
\vspace{-0.6cm}
\begin{equation} \label{equ:opt_problem_hydraulics}
\begin{array}{llr}
{\mbox{find}} & \m{q} \\
\mbox{subject to} & \m F_{\mr d} \m q - \m d = \m 0, & ~~~~~~\textit{Mass Conservation}\\ 
& \m F_{\mr l} \m q = \m 0, & ~~~~~~~~~~~~\textit{Energy Conservation}\\ 
& \m q \in \mathbb{R}^n,~\m F_d \in \mathbb{R}^n \times \mathbb{R}^{n_{\mr{p}}},~\m F_l \in \mathbb{R}^n \times \mathbb{R}^{n_l}, \m d \in \mathbb{R}^n\\
\end{array}
\end{equation}
\end{theo}
and determines the flow discharges such that the energy and mass balance are satisfied. This problem is non-convex, non-smooth, and non-linear due to the friction losses.

\paragraph{WDN Design Optimization Modeling (DOM)}
The decision variables for this problem are the flows $\m q$, the tank heads $\m h_{\mr{r}}$, and the height of the tanks ($\m h_{\mr b}$). All pipes are constrained by a minimum ($\m p_{\mr{min}}$) and maximum ($\m p_{\mr{max}}$) pressure. For the cases of designing the pipe diameters altogether while satisfying mass and energy constraints, a new constraint in the flow velocities would also be required to allow minimum and maximum values. This problem, however, is untractable to solve with simple gradient solvers available in Excel and is hence outside of the scope of this paper. The objective function of the problem is the network net present value that considers the pipeline costs, the pump energy costs, and the tank costs such that the network can deliver acceptable flow pressures while having a reduced overall cost. This problem is stated in Eq.~\eqref{equ:design_optimization}.

\begin{theo}[\textbf{WDN Design Optimization Modeling}] 
\vspace{-0.6cm}
\begin{equation} \label{equ:design_optimization}
\begin{array}{ll}
\underset{\m q, \m h_{\mr{r}},\m z_{\mr b}}{\mbox{minimize}} & \text{Eq}.~\eqref{equ:overall_cost} \\
\mbox{subject to} & \m F_{\mr d} \m q - \m d = \m 0,  ~~~~~~\textit{Mass Conservation}\\ 
& \m F_{\mr l} \m q = \m 0,  ~~~~~~~~~~~~\textit{Energy Conservation}\\ 
& \m p_{\mr{min}} \leq \m p \leq \m p_{\mr{max}} \\
& \m h_{r_{\mr{min}}} \leq \m h_{\mr{r}} \leq \m h_{r_{\mr{max}}} \\
& \m 0 \leq \m z_{\mr b} + \m h_{\mr{r}} - \m z_{\mr r} \leq \m z_{\mr{max}} \\
& \m q \in \mathbb{R}^n, \m d \in \mathbb{R}^n, \m q \in \mathbb{R}^{n_{\mr d}}, 
\m p \in \mathbb{R}^{n_{\mr d}}, 
\m h_{\mr r} \in \mathbb{R}^{n_{\mr r}}, 
\m z_{\mr b} \in \mathbb{R}^{n_{\mr r}}, \\ 
& \m F_d \in \mathbb{R}^n \times \mathbb{R}^{n_{\mr{p}}}, \m F_l \in \mathbb{R}^n \times \mathbb{R}^{n_l},\\

\end{array}
\end{equation}
\end{theo}

\subsection{Excel Framework and data entry}
\paragraph{Excel WDN Sheet - X-WHAT.xlsx}
In this section, we describe the Looped Hydraulic Network simple Excel sheet used to solve WDN problems using Excel solvers. Only Microsoft Excel is required to run the model; although other solvers can be coupled with the network model, such as Open Solver \citep{OpenSolver}. Users can define solver properties such as using Multi-Start points to test the objective function from multiple initial points of the decision space, hence increasing the chances, although not guaranteeing, of finding the global optima. In addition, users can define a different cost function that can represent other quantities that are not monetary costs, such as the average pressure in the network.

\begin{figure}
    \centering
    \includegraphics[scale = 0.28]{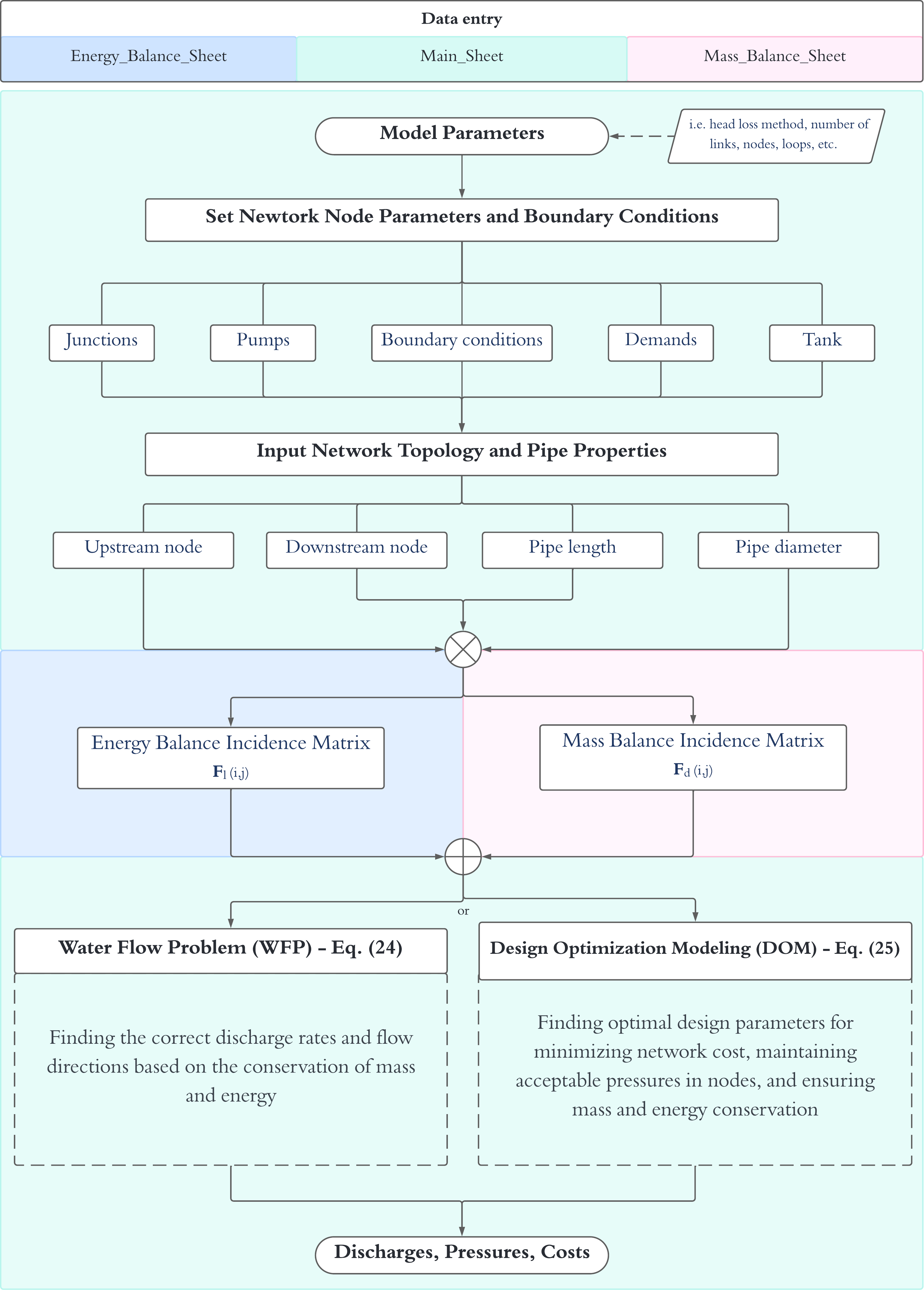}
    \caption{General data entry flowchart. No macros or VBA coding are required to run the model. Boundary conditions can be specified for nodes where the pressure is known.}
    \label{fig:user_algorithm}
\end{figure}

The data entry procedure is systematic and involves three sheets: Main, Mass Balance, and Energy Balance. Fig.~\ref{fig:user_algorithm} outlines the general steps for entering all necessary information. The process can be divided into five main sections, with the first section on the main sheet dedicated to setting the model parameters listed in Tab.~\ref{tab:model_parameters}. Note that it is not mandatory to fill in all the parameters listed.  The essential parameters vary depending on the analysis to be performed.  For example, for a basic hydraulic simulation that is typically addressed in a classroom environment, only the first six parameters are required.


\begin{table}[]
\centering
\resizebox{\textwidth}{!}{%
\begin{tabular}{@{}clcc@{}}
\toprule
Required Parameters & \multicolumn{3}{c}{Description}                                                                                                            \\ \midrule
D-W or H-W          & Head loss method (H-W: Hazen-Williams; D-W: Darcy-Weisbach)                         & \multirow{6}{*}{\rotatebox{90}{WFP}}         & \multirow{17}{*}{\rotatebox{90}{DOM}} \\
$\nu$                   & Represents the kinematic viscosity of the fluid in $\mr{m^2 \cdot s^{-1}}$                           &                              &                       \\
$\rho$                   & Represents the density of the fluid, measured in kg/m³.                             &                              &                       \\
Links               & Indicates the total number of links within the hydraulic network.                   &                              &                       \\
Nodes               & Refers to the total number of nodes or junction points within the network.          &                              &                       \\
Loops               & Indicates the total number of loops within the hydraulic network.                   &                              &                       \\ \cmidrule(r){1-3}
$P_c$                  & \multicolumn{2}{l}{Stands for the cost per kilowatt-hour (kWh) of energy in USD/kWh.}                              &                       \\
$i_r$                  & \multicolumn{2}{l}{Represents the annual increase rate in energy costs, expressed as a percentage.}                &                       \\
Years               & \multicolumn{2}{l}{Indicates the lifespan of the system being analyzed, measured in years.}                        &                       \\
Rate          & \multicolumn{2}{l}{Refers to the interest rate applicable to investments or loans related to the project, expressed as a percentage.}                  &  \\
$\alpha$             & \multicolumn{2}{l}{Represents the operational cost rate associated with maintenance and operation expenses over time; it’s expressed as a percentage.} &  \\
$k_1$                  & \multicolumn{2}{l}{The day factor used for calculations related to daily operations or impacts within the system.} &                       \\
$k_2$                  & \multicolumn{2}{l}{The hour factor, similar to k1 but applied on an hourly basis for more granular analysis.}      &                       \\
$D_{tr}$           & \multicolumn{2}{l}{Duration to fill the reservoir, indicating the time required (in days) to completely fill up storage reservoirs within the system.} &  \\
Material Cost & \multicolumn{2}{l}{Refers to the cost per cubic meter (USD/m³) associated with materials required for construction of the reservoirs.}                 &  \\
$V_k$            & \multicolumn{2}{l}{Wind speed velocity (m/s) used to calculate mechanical stresses on structures exposed above the surface level (reservoirs).}        &  \\
$p$             & \multicolumn{2}{l}{p is an exponent that increases with the topographic elevation and can be estimated in terms of the wind speed}                     &  \\ \bottomrule
\end{tabular}%
}
\caption{Model parameters required according to the type of analysis to be performed.}
\label{tab:model_parameters}
\end{table}

In the second section, the network node parameters and boundary conditions are defined. Each node must have its ground elevation specified, along with the identification of whether it is a manhole or a tank. Pumps are included in the analysis by entering information such as daily operating hours, efficiency, ground elevation, and linear head loss coefficient. Tank specifications, including elevation and height above ground, must be provided. However, the height above the ground can also be included as a decision variable in the optimization design modeling (DOM). Furthermore, demands for each network node are defined by specifying whether flows enter or leave a node. Boundary conditions are defined specifying whether a node has a fixed head or not. When the type of node is designated as a tank, it requires a fixed head.

Following the specification of the node parameters, the third section should be used to enter detailed information on the links. For each link in the network, the upstream and downstream nodes must be specified, as well as the pipe length and diameter. 

Sections four and five involve completing the mass and energy balance incidence matrices. The mass-balance sheet includes the mass balance incidence matrix ($F_d$) for carrying out the mass balance calculation, while the energy-balance sheet requires completion of the energy balance incidence matrix ($F_l$). The order of completion is not mandatory. More details of how to enter the input data in the file are presented in the supplemental material.

After completing the data entry process through the Main, Mass-Balance, and Energy-Balance sheets, users can proceed to perform the desired analysis using the Excel Solver. In WFP, the objective is to determine the correct discharges and flow directions based on the principles of mass and energy conservation as described in Eq.~\eqref{equ:opt_problem_hydraulics}. Users may also choose a DOM approach which usually involves minimizing the cost function as described in Eq.~\eqref{equ:design_optimization}. However, different analyses can be performed depending on the design parameters to be optimized. 

Once the analysis is complete, the results are presented in both tabular and graphical formats. Additional information on potential analyses can be found in Section 3 - Numerical Case Studies.

\subsubsection{Mean Average Error Metric}
To  compare the modeling results from the spreadsheet with EPANET or with available solutions in published literature, we use the mean absolute error (MAE), written as:

\begin{equation} \label{equ:MAE}
    \mr{MAE} = \frac{\sum_{i=1}^n\left|x_{\mr m}-x_{\mr{obs}}\right|}{n}
\end{equation}
where $n$ is the number of observations, $x_{\mr m}$ is the modeled state and $x_{\mr{obs}}$ is the observed or truth-based state. This metric is applied for head pressures in all nodes and flow discharges in all links and the units are the same as the state evaluated.

\section{Numerical Case Studies} \label{sec:case_studies}
 In this section we detail the case studies used to test the model developed. We choose three different testing cases (a), (b), and (c), with increasing complexity. For all tested cases, a lifespan ($n_\mr{l}$) of 25 years, $i_{\mr{r}} = 12\%$, and $i_{\mr e} = 6\%$ is assumed. The pipeline cost function can be approximated with 7-order polynomial coefficients of \citep{cassiolato2023minlp}:

$$C_l(D) = 2.27\times 10^{-13}D^7 - 3.32\times 10^{-10}D^6 + 1.85\times 10^{-7}D^5 - 4.74\times 10^{-5}D^4 + 5.76\times 10^{-3}D^3 - 1.87\times 10^{-1}D^2 + 4.1$$
with $D$ in mm, which is a continuous function with derivatives calculated and later used in the gradient solver for the optimization problem. The pump elevation $z_{\mr p}$ is assumed 5 m below the tank ground elevation $z_{\mr r}$.

The tank material cost is assumed as 60 $\mr{USD\cdot m^{-2}}$.

\subsection{Numerical Case Study 1: Water Flow Problem of Eq.~\eqref{equ:opt_problem_hydraulics}}
All problems presented in this numerical case study are only simulated to estimate flow discharges and flow directions. The values of the tank water depths are manually designed (i.e., by trial and error) to be the minimum value to allow for a pressure head in all nodes within 10 to 30 m. The ground elevations of the tanks are the same as those of the node to which they are connected. For this analysis, all tanks are leaned on the ground (i.e., $h_b = 0$). This is typically the case for tanks with relatively large volumes. The lengths of the pumps to the tanks are fixed in 500 m and the pumping flow discharges are calculated using Eq.~\eqref{equ:pump_flow}.

The testing cases presented in the following subsection are representative of different scales. The first is a small rural community with a population of 160 inhabitants. The second represents a larger network that could accommodate a population of 12,800
 inhabitants. Finally, the last case study is a city with a population of 44,416 inhabitants. The rationale for simulating varied cases is to test the modeling capacity for different demands, tank volumes, and increase the reliability of the model.

\subsubsection{Testing Case (a) - The 2-loop, 5-nodes, 1-tank network}
To illustrate the numerical procedures to solve a water distribution network problem using a simple spreadsheet, in this case study, we develop a step-by-step mathematical definition of the problem and solve it numerically in Excel. In addition, we compare the results obtained by the simple spreadsheet with the EPANET solver.

The network of this case study has 5 nodes, with pipe segments of 100 m and Hazzen Willians coefficients of 130. The network is supplied by an upstream tank that has a fixed-head boundary condition of 20.84 meters above the ground elevation. In this case, all pipes are fixed with a 40 mm diameter. The pipeline cost is USD 1,949.24. The tank volume is 17.28 $\mr{m^3}$.

The problem consists in defining the flow directions and discharges in the network such that the conservation of energy and mass is satisfied for all loops and nodes.

\begin{figure}
    \centering
    \includegraphics[scale = 0.15]{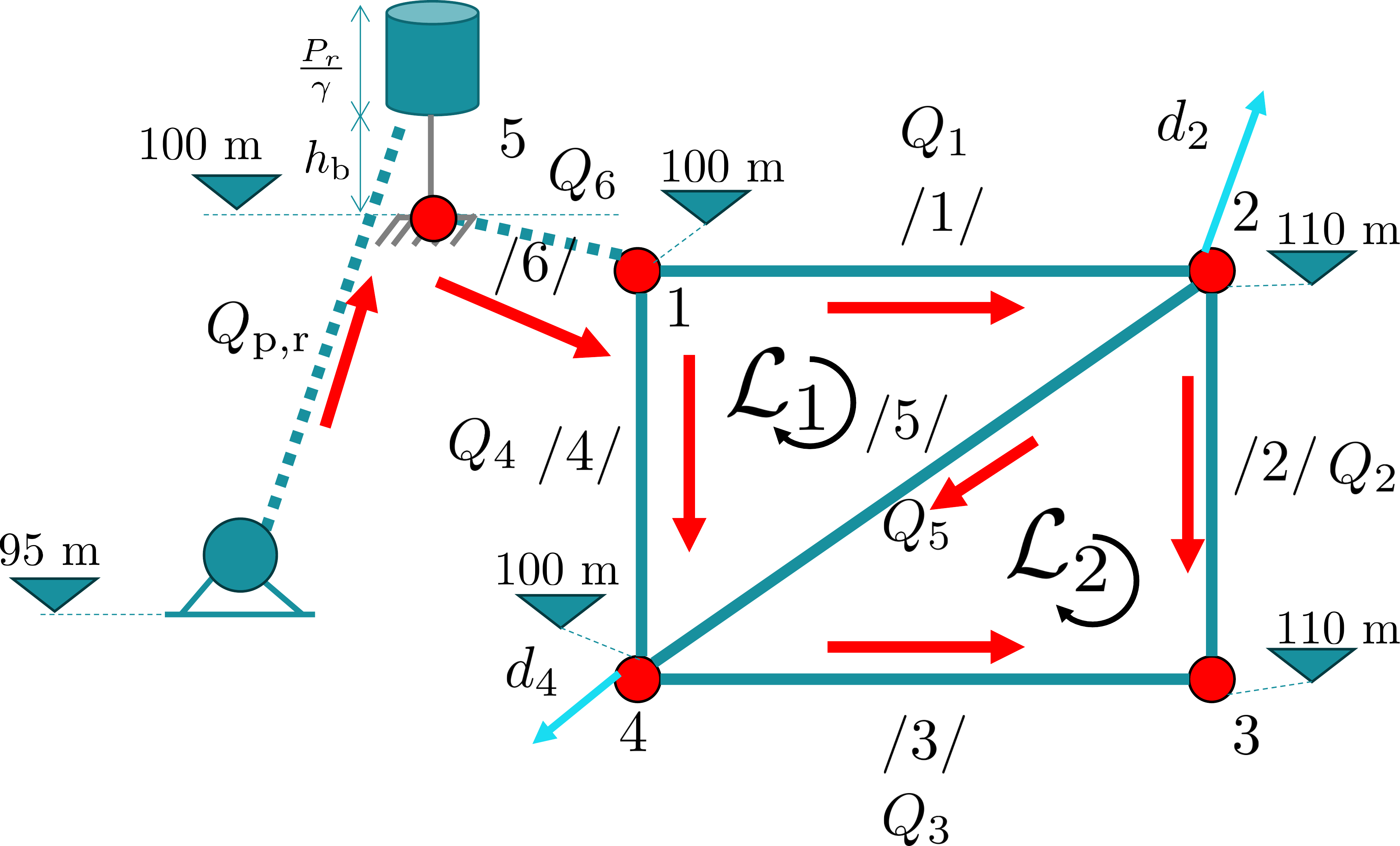}
    \caption{Network of Numerical Case Study 1, testing case (a), where $P_{\mr{r}}/\gamma$ is the tank water depth and the tank head is given by $z_{\mr{r}} + P_{\mr{r}}/\gamma.$ The tank is elevated $h_{\mr b}$ from the ground elevation. Nodes 2 and 3 are 10 m above all other nodes, increasing the required energy to allow sufficient dynamic pressure at them.}
    \label{fig:enter-label}
\end{figure}

\begin{table}
\centering
\begin{tblr}{
  hline{1-2,7} = {-}{},
}
Node & Ground Elevation [m] & Demand [L/s] & $P_r/\gamma$ \\  
1    & 100   & 0  & -         \\            
2    & 110   & 0.2   & -            \\           
3    & 110   & 0        & -             \\        
4    & 100   & 0.3                  \\       
5    & 100      & -0.5     & 20.84                    
\end{tblr}
\caption{Node data for numerical case study 1 (a). The pressure head in the tank was determined by trial and error aiming to find the minimum pressure, with the tank leaned in the ground, that satisfy a pressure within 10 and 30 $\mr{mH_2O}$ in all nodes.}
\label{tab:node_data}
\end{table}

\begin{table}
\centering
\caption{Link data for numerical case study 1.}
\begin{tblr}{
  hline{1-2,8} = {-}{},
}
Link ID & {Upstream\\ Node} & {Downstream\\ Node} & L [m] & C   & D [mm] & $Q_{m}$ [L/s]     & $Q_{obs}$ [L/s]   \\
/1/     & 1                 & 2                   & 100   & 130 & 40     & 0.249  & 0.250  \\
/2/     & 2                 & 3                   & 100   & 130 & 40     & 0.020  & 0.250  \\
/3/     & 4                 & 3                   & 100   & 130 & 40     & -0.020 & 0.020  \\
/4/     & 1                 & 4                   & 100   & 130 & 40     & 0.251  & -0.020 \\
/5/     & 2                 & 4                   & 100   & 130 & 40     & 0.029  & 0.030  \\
/6/     & 5                 & 1                   & 100   & 130 & 40     & 0.500  & 0.250  
\end{tblr}
\caption{Input data of testing case (a) and modeled and observed flow discharges at the links.}
\label{tab:testing_case_a}
\end{table}

The flow direction matrix that for each node (cols 1 to 5) matches the links that are associated with by the columns from 1 to 6 results in:

$$
\m F_d = \left[\begin{array}{cccccc}
-1 & 0 & 0 & -1 & 0 & 1 \\
1 & -1 & 0 & 0 & -1 & 0 \\
0 & 1 & 1 & 0 & 0 & 0 \\
0 & 0 & -1 & 1 & 1 & 0 \\
0 & 0 & 0 & 0 & 0 & -1.
\end{array}\right]
 $$ 
The incidence loop matrix that associate every loop (from 1 to 2 in the lines) with every pipe (from columns 1 to:
$$
\m F_l = \left[\begin{array}{ccccc}
1 & 0 & 0 & -1 & 1 \\
0 & 1 & -1 & 0 & -1
\end{array}\right]
$$

All pipes have the same length (100 m) and Hazzen-Willians coefficient, resulting in:

$$
k \Big|_\text{H-W} = \frac{10.67}{130^{1.85}[(40/1000)/100]^{4.87}} = 842,048.4
~\mr{m^{-0.85}},
$$

Therefore, the mass balance and energy constraints are written as:

$$
\begin{aligned}
& {\left[\begin{array}{cccccc}
-1 & 0 & 0 & -1 & 0 & 1 \\
1 & -1 & 0 & 0 & -1 & 0 \\
0 & 1 & 1 & 0 & 0 & 0 \\
0 & 0 & -1 & 1 & 1 & 0 \\
0 & 0 & 0 & 0 & 0 & -1
\end{array}\right]\left[\begin{array}{l}
Q_1 \\
Q_2 \\
Q_3 \\
Q_4 \\
Q_5
\end{array}\right]-\left[\begin{array}{c}
0 \\
0.2 \\
0 \\
0.3 \\
-0.5
\end{array}\right]=\left[\begin{array}{l}
0 \\
0 \\
0 \\
0 \\
0
\end{array}\right]} \\
& {\left[\begin{array}{ccccc}
1 & 0 & 0 & -1 & 1 \\
0 & 1 & -1 & 0 & -1
\end{array}\right]\left[\begin{array}{l}
842,048.4 Q_1\left|Q_1\right|^{0.85} \\
842,048.4 Q_2\left|Q_2\right|^{0.85} \\
842,048.4 Q_3\left|Q_3\right|^{0.85} \\
842,048.4 Q_4\left|Q_4\right|^{0.85} \\
842,048.4 Q_5\left|Q_5\right|^{0.85}
\end{array}\right]=\left[\begin{array}{l}
0 \\
0
\end{array}\right]}
\end{aligned}
$$
and can be applied in the WDN purely hydraulic network model from Eq.~\eqref{equ:opt_problem_hydraulics}, resulting in a non-convex, non-linear and non-smooth problem due to the friction modeling, although the mass balance equations are linear. Students can derive the aforementioned system of equations in class without excessive effort and the techniques to solve the problem can be discussed. To solve the problem in the Excel, we use the GRG non-linear \citep{smith1992solving} solver, which is a gradient based solver that has shown acceptable results for relatively small networks.

\subsubsection{Testing Case (b) - Porto (2004) Network}
\begin{figure}
    \centering
    \includegraphics[scale = 0.14]{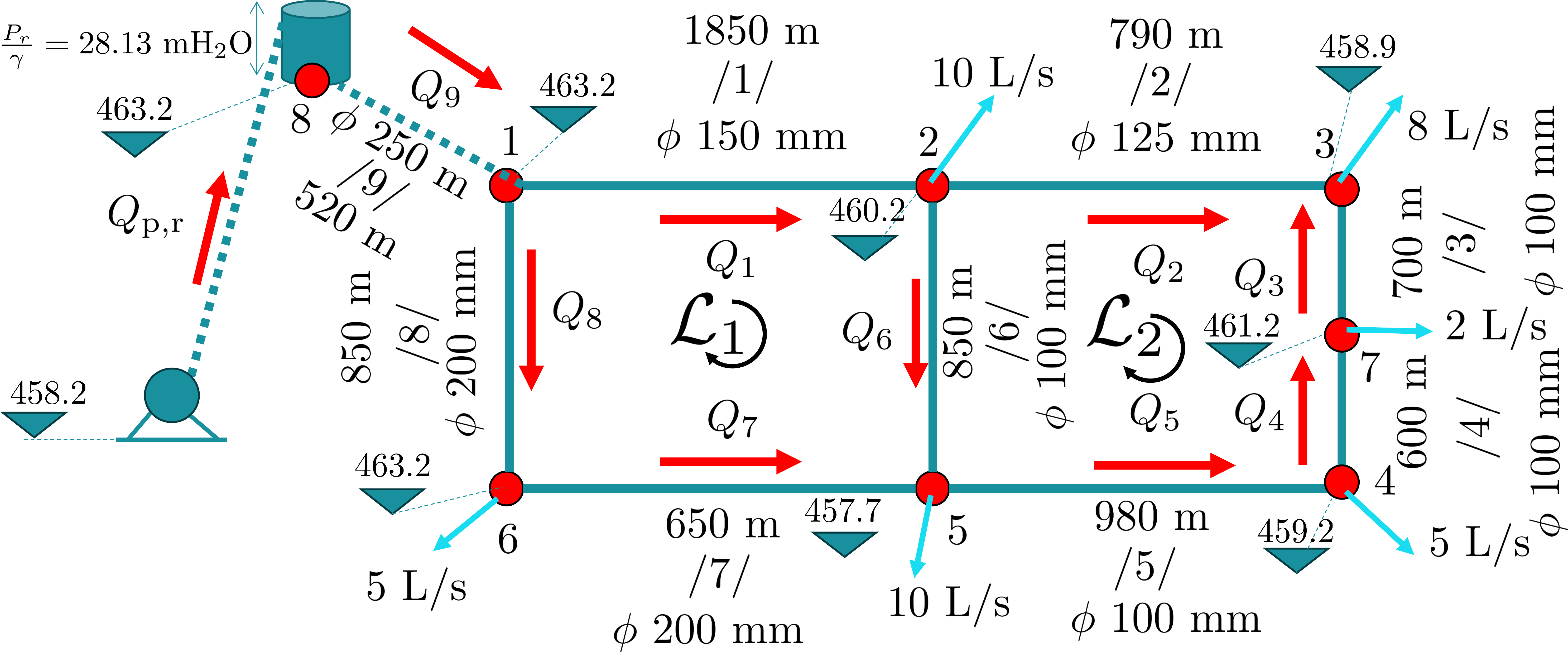}
    \caption{Schematic of the testing case (b), adapted from \citep{porto2004hidraulica}.}
    \label{fig:porto_network}
\end{figure}

\begin{figure}
    \centering
    \includegraphics[scale = 0.17]{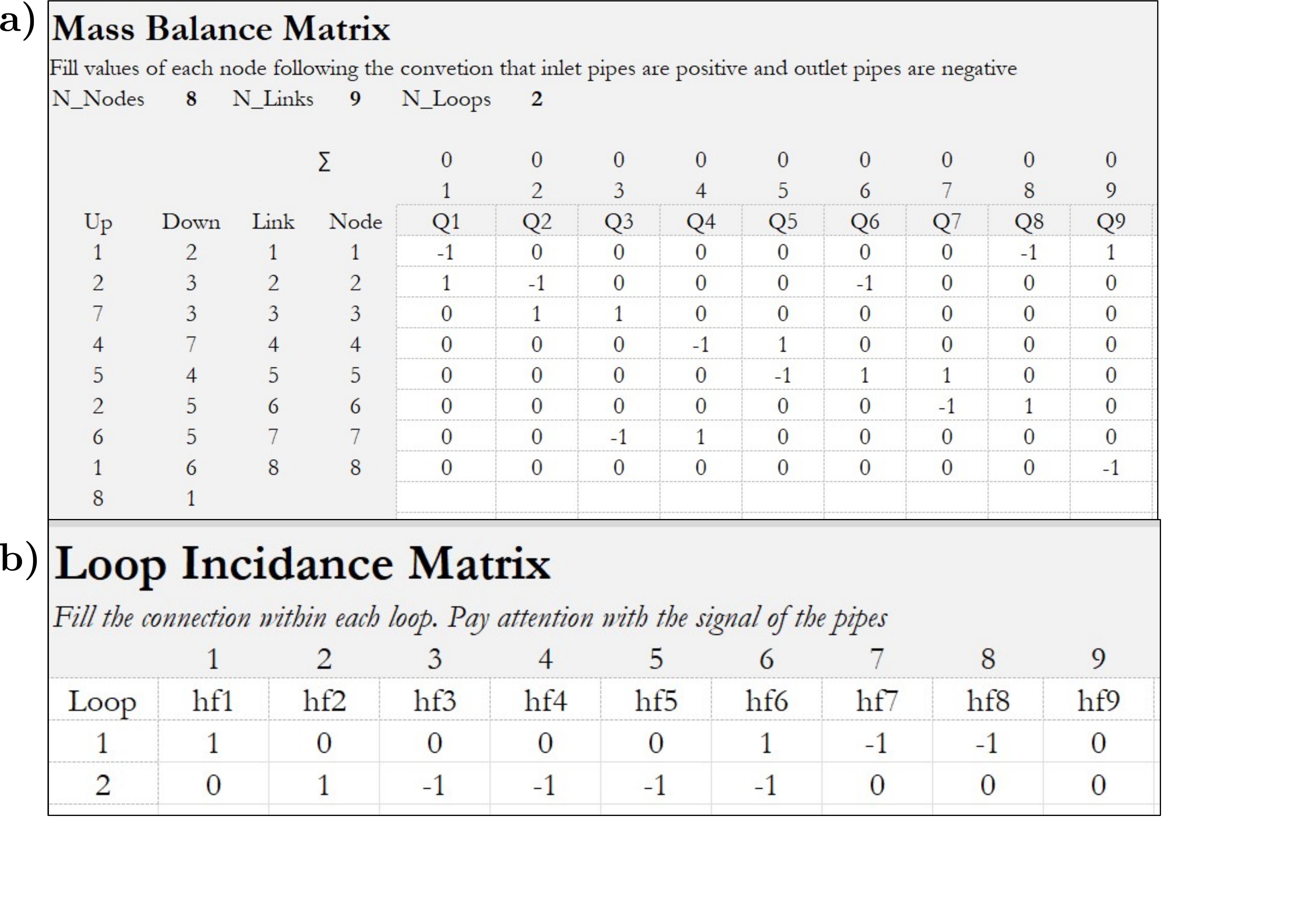}
    \caption{Incidence matrices used to simulate the testing case (b). Part (a) is the flow direction matrix and (b) is the loop incidence matrix. Each matrix is entered in a different Excel sheet.}
    \label{fig:Porto_Matrices}
\end{figure}

This case study is a 2-loop network supplied by a tank adapted from \citep{porto2004hidraulica}. All pipes have rugosity of 0.0015 mm and the friction losses are hence modeled with the Darcy-Weisbach model. The kinematic viscosity of the water is assumed to be $10^{-6}~\mr{m^2 \cdot s^{-1}}$. The input data for the nodes are presented in Tab.~\ref{tab:num_cas_study2_nodes}. The pressure head is $16.11~\mathrm{mH_2O}$. The link input data are presented in Tab.~\ref{tab:porto_links}. The schematic of the problem is presented in Fig.~\ref{fig:porto_network}. To illustrate the entry of matrices $\m F_d$ and $\m F_l$ in the developed Excel spreadsheet, Fig.~\ref{fig:Porto_Matrices} is presented. The network pipeline cost is 126,509.50 USD. The tank volume is 1,382.4 $\mr{m^3}$.

\begin{table}
\centering
\begin{tblr}{
  hline{1-2,10} = {-}{},
}
Node & Ground Elevation [m] & Demand [L/s] & $P/\gamma$ [mH2O] \\
1    & 463.20        & 0         & -              \\
2    & 460.20        & 10        & -              \\
3    & 468.90        & 8        & -              \\
4    & 471.20        & 5        & -              \\
5    & 467.70        & 10        & -              \\
6    & 463.20        & 5         & -              \\
7    & 459.20        & 2         & -              \\
8    & 463.20        & -40       & 28.7          
\end{tblr}
\caption{Node data of Numerical Case Study 1 for the Porto (2004) network.}
\label{tab:num_cas_study2_nodes}
\end{table}

\begin{table}
\centering
\caption{Link input data for the testing case (b) \citep{porto2004hidraulica}.}
\label{tab:porto_links}
\begin{tblr}{
  hline{1-2,11} = {-}{},
}
Link ID & Upstream Node & Downstream Node & L [m] & D [mm] & $\epsilon$ [mm] & $Q_{\mr{m}}$ [L/s] & $Q_{\mr{obs}}$ [L/s] \\
/1/     & 1             & 2               & 1850  & 150    & 0.0015          & 13.97          & 13.98                                            \\
/2/     & 2             & 3               & 790   & 125    & 0.0015          & 8.72           & 8.72                                             \\
/3/     & 7             & 3               & 700   & 100    & 0.0015          & -0.72          & -0.72                                            \\
/4/     & 4             & 7               & 600   & 100    & 0.0015          & 1.28           & 1.28                                             \\
/5/     & 5             & 4               & 980   & 100    & 0.0015          & 6.28           & 6.28                                             \\
/6/     & 2             & 5               & 850   & 100    & 0.0015          & -4.74          & -4.74                                            \\
/7/     & 6             & 5               & 650   & 200    & 0.0015          & 21.03          & 21.02                                            \\
/8/     & 1             & 6               & 850   & 200    & 0.0015          & 26.03          & 26.02                                            \\
/9/     & 8             & 1               & 520   & 250    & 0.0015          & 40.00          & 40.00                                            
\end{tblr}
\end{table}

\subsubsection{Testing Case (c) - Huddlestone (2004) Network}
\begin{figure}
    \centering
    \includegraphics[scale = 0.15]{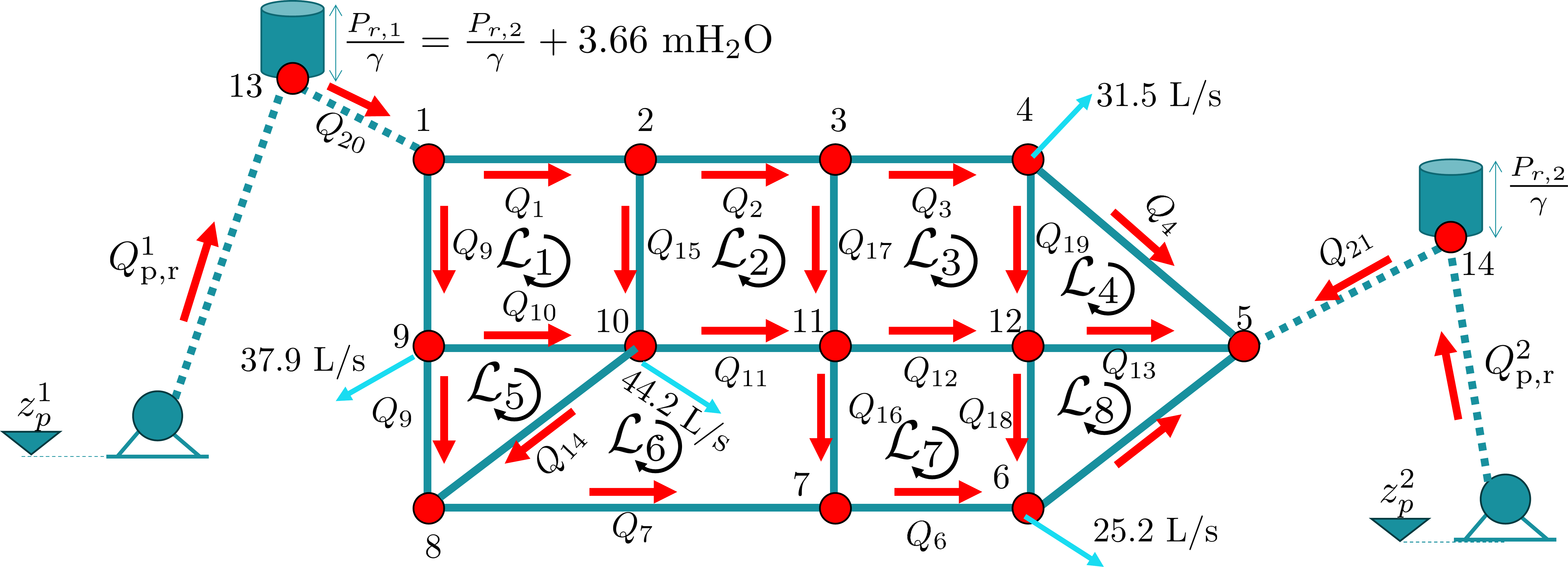}
    \caption{Network schematics of testing case (c), adapted from \citep{huddleston_water_2004}.}
    \label{fig:Hud_network}
\end{figure}

This numerical case study shows a larger network with 8 loops, 12 internal nodes, 2 tanks, and a total of 21 links \citep{huddleston2004water}. All nodes have the same elevation. Tank 1 (node 13) is $\mr{3.66}~\mr{mH_2O}$ above Tank 2 (node 14) (see Fig.~\ref{fig:Hud_network}). Demands are taken in four nodes of the network and the problem consists of finding the discharges and flow directions in all pipes subject to known tank head boundary conditions. The topological matrices $\m F_d$ and $\m F_l$ of this problem are shown in the Supplementary Material. The network pipeline cost is USD 2,093,182.81 USD. The pipe information is given in Tab.~\ref{tab:hud_table}. The tank 1 and 2 volumes are 3,572.03 and 1,224.89, $\mr{m^3}$ respectively.

\begin{table}
\centering
\begin{tabular}{ccccccc}
\hline \begin{tabular}{l} 
Link \\
ID
\end{tabular} & \begin{tabular}{l} 
$D$ \\
$[\mathrm{mm}]$
\end{tabular} & \begin{tabular}{l} 
$L$ \\
$[\mathrm{m}]$
\end{tabular} & \begin{tabular}{l} 
$\epsilon$
$(\mathrm{mm})$
\end{tabular} & \begin{tabular}{l} 
$Q_h$\\
 $\left(L / \mathrm{s}\right)$
\end{tabular} & \begin{tabular}{l} 
$Q_m$ \\
 $\left(L / \mathrm{s}\right)$
\end{tabular} \\
\hline
/1/  & 305  & 457.2 & 0.26 & 55.8  & 55.5  \\
/2/  & 203  & 304.8 & 0.26 & 40.0  & 39.8  \\
/3/  & 203  & 365.8 & 0.26 & 16.5  & 16.4 \\ 
/4/  & 203  & 609.6 & 0.26 & -10.3 & -10.4 \\
/5/  & 203  & 853.4 & 0.26 & -8.7  & -8.7  \\
/6/  & 203  & 335.3 & 0.26 & 12.6  & 12.5  \\
/7/  & 203  & 304.8 & 0.26 & 15.0  & 14.9  \\
/8/  & 203  & 762   & 0.26 & 9.7   & 9.6   \\
/9/  & 203  & 243.8 & 0.26 & 48.0  & 47.8  \\
/10/ & 152  & 396.2 & 0.26 & 0.4   & 0.3   \\
/11/ & 152  & 304.8 & 0.26 & 10.8  & 10.7  \\
/12/ & 254  & 335.3 & 0.26 & -7.4  & -7.6  \\
/13/ & 254  & 304.8 & 0.26 & -16.0 & -16.3 \\
/14/ & 152  & 548.6 & 0.26 & 5.3   & 5.3   \\
/15/ & 152  & 335.3 & 0.26 & 15.7  & 15.7  \\
/16/ & 152  & 548.6 & 0.26 & -2.4  & -2.4  \\
/17/ & 254  & 365.9 & 0.26 & 23.6  & 23.5  \\
/18/ & 152  & 548.6 & 0.26 & 4.0   & 4.0   \\
/19/ & 152  & 396.2 & 0.26 & -4.7  & -4.7  \\
/20/ & 1000 & 25    & 0.26 & 103.7 & 103.4 \\
/21/ & 1000 & 25    & 0.26 & 35.1  & 35.4  \\
\hline
\end{tabular}
\caption{Input data for Huddlestone Network (first 4 columns), where $Q_h$ is the modeled results from the paper and $Q_m$ are the results with the modeled developed in this paper.}
\label{tab:hud_table}
\end{table}

\subsection{Numerical Case Study 2 - Design Optimization using Eq.~\eqref{equ:design_optimization}}
We solve the design optimizatin problem using the network presented in testing cases cases (a), (b), and (c) from Numerical Case Study 1, but now optimizing the tank pressure head and height from the ground. In all testing cases in this numerical case study, pressure heads in the internal nodes of the network are ensured to be higher than $10~\mr{mH_2O}$ and lower than $30~\mr{mH_2O}$. 

In addition, tanks have a minimum and maximum static pressure of 0.25 and 40~m. The tank height is limited by a minimum of 0 and a maximum of 39.5 m. The pipelines connecting the pumps to the tanks are 500 m in length and their diameters are designed not to allow flow velocities greater than $2\mr{m\cdot s^{-1}}$, hence finding the closest diameter that satisfies this condition. The values of $k$ are calculated using the headloss model of each testing case (i.e., using H-W \eqref{equ:kHW} for (a), and D-W \eqref{equ:kDW} for (b) and (c)). The pump discharge, the diameter of the supply link, and the friction coefficient are shown in Tab.~\ref{tab:pump_table} for all the testing cases. All pumps operate 12 h per day ($n_h = 12$) with an average efficiency of 85\% ($\eta_p$). The Hazzen-Willians coefficient or the rugosity of the Darcy-Weisbach model is assumed to be the same as the WDN simulated. The interest rate is 12\% ($i_r$), the annual increase in energy costs is 6\% ($i_e$) and the lifespan of the WDN is 25 years.

\begin{table}
\centering
\begin{tblr}{
}
\hline \\
Case Study & Tank & $Q_{\mr{p,r}}$ [L/s] & $D$ [mm] & $k^p$      \\
\hline
(a)          & 1         & 133.3     & 40    & 4,210,242.04 \\
(b)          & 1         & 53.33      & 200    & 3655.84 \\
(c)          & 1         & 137.81     & 300    & 336.78  \\
(c)          & 2         & 47.26      & 175    & 5641.59 \\
\hline
\end{tblr}
\caption{Input data used to simulate the head losses in the pumping to the tank.}
\label{tab:pump_table}
\end{table}

\section{Results} \label{sec:results}

In this section, we summarize the results of the numerical case studies presented in this paper. It is important to note that we have included and listed detailed results, such as friction factors, Reynolds numbers, head loss coefficients, piezometric heads, and pressure heads, for all numerical testing cases in the supplemental material of the paper. This decision was made to maintain brevity in the main text and to focus on the major outcomes.


\subsection{Numerical Case Study 1}
\subsubsection{Testing Case (a) - The 2-loop, 5-nodes, 1-tank network}
The network hydraulic simulation results are presented in Tab.~\ref{tab:testing_case_a}. The MAEs for flows and pressures are, in comparison with EPANET, 0.074 $\mr{L\cdot s^{-1}}$ and 0.016 $\mr{m}$, respectively. 

\subsubsection{Testing Case (b) - Porto (2004)}
The results of the solution of the problem in Eq.~\eqref{equ:opt_problem_hydraulics} are presented in Tab.~\ref{tab:porto_links}. The MAEs of flows and pressures, compared to EPANET, are 0.004 $\mr{L\cdot s^{-1}}$ and 0.058 $\mr m$, respectively.

\subsubsection{Testing Case (c) - Huddlestone (2004) Network}
The results of the network Water Flow Problem are presented in Tab.~\ref{tab:hud_table}. The MAE of the flow simulated compared to the ones reported in \citep{huddleston_water_2004} is 0.13 $\mr L \cdot \mr s^{-1}$, less than 0.1\% of the maximum flow.

\subsection{Numerical Case Study 2}
The foundation states modeled for the optimization design problem of testing cases (a), (b), and (c) are presented in Tab.~\ref{tab:foundation_states}. The pump states are shown in Tab.~\ref{tab:pump_states}. The overall cost composition including foundation, tank material, and pump energy costs is presented in Tab.~\ref{tab:overall_costs}.

\begin{table}
\centering
\begin{tblr}{
  cell{1}{1} = {r=2}{},
  cell{1}{2} = {c=2}{},
  cell{1}{4} = {c=2}{},
  cell{1}{6} = {c=2}{},
  cell{1}{8} = {c=2}{},
  cell{1}{10} = {c=2}{},
  hline{1,3,7} = {-}{},
}
Case
  Study & D [m] &       & $h_{b}$ [m] &      & $h_r$ [m] &       & $H_k$ [kN] &        & $M_k$ [$\mr{kN \cdot m}$] &         \\
             & DOM   & WFP   & DOM    & WFP  & DOM    & WFP   & DOM     & WFP    & DOM       & WFP     \\
(a) - R1     & 2.15  & 1.03  & 16.07  & 0.00 & 4.77   & 20.84 & 10.67   & 17.79  & 197.14    & 209.57  \\
(b) - R1     & 7.83  & 7.83  & 0.00   & 0.00 & 28.70  & 28.70 & 205.56  & 205.57 & 3334.45   & 3334.68 \\
(c) - R1     & 14.65 & 18.03 & 0.00   & 0.00 & 21.19  & 13.99 & 259.27  & 185.97 & 3105.64   & 1470.55 \\
(c) - R2     & 9.43  & 12.29 & 0.00   & 0.00 & 17.53  & 10.33 & 130.46  & 85.44  & 1292.82   & 498.86  
\end{tblr}
\caption{Foundation states for all case studies simulated, where DOM = Design Optimization Modeling [Eq.~\eqref{equ:design_optimization}] and WFP = Water Flow Problem [Eq.~\eqref{equ:opt_problem_hydraulics}].}
\label{tab:foundation_states}
\end{table}

\begin{table}
\centering
\begin{tblr}{
  cell{1}{1} = {r=2}{},
  cell{1}{2} = {c=2}{},
  cell{1}{4} = {c=2}{},
  cell{1}{6} = {c=2}{},
  hline{1,3,7} = {-}{},
}
Case
  Study & $h_g$ [m] &       & $H_{\mr{m,p}}$ [m] &       & $P$ [kW] &       \\
             & DOM    & WFP   & DOM      & WFP   & DOM      & WFP   \\
(a) - R1     & 25.84  & 25.84 & 31.44    & 31.44 & 0.24     & 0.24  \\
(b) - R1     & 33.70  & 33.70 & 44.10    & 44.10 & 27.14    & 27.14 \\
(c) - R1     & 26.19  & 18.99 & 32.59    & 25.39 & 51.83    & 40.38 \\
(c) - R2     & 22.53  & 15.33 & 35.13    & 27.93 & 19.16    & 15.23 
\end{tblr}
\caption{Pump states for all testing cases, where DOM = Design Optimization Modeling [Eq.~\eqref{equ:design_optimization}] and WFP = Water Flow Problem [Eq.~\eqref{equ:opt_problem_hydraulics}].}
\label{tab:pump_states}
\end{table}

\begin{table}
\small
\centering
\begin{tblr}{
  cell{1}{1} = {r=2}{},
  cell{1}{2} = {c=2}{},
  cell{1}{4} = {c=2}{},
  cell{1}{6} = {c=2}{},
  cell{1}{8} = {c=2}{},
  cell{1}{10} = {c=2}{},
  hline{1,3,6} = {-}{},
}
Testing
  Case & $C_{\mr p}$ [$10^3$USD] &        & $C_{\mr m}$ [$10^3$USD] &        & $C_{\mr f}$ [$10^3$USD] &        & $C_{\mr t} - C_{\mr p}$ [$10^3$USD] &         & $C_{\mr t}$ [$10^3$USD] &         \\
             & DOM          & WFP    & DOM            & WFP    & DOM              & WFP    & DOM           & WFP     & DOM       & WFP     \\
(a) & 0.53  & 0.53  & 2.86   & 4.14   & 1.52   & 1.74   & 10.99   & 12.48   & 12.93   & 14.43    \\
(b) & 23.78 & 23.78 & 87.63  & 87.63  & 55.43  & 55.43  & 439.31  & 439.31  & 565.81  & 565.82   \\
(c) & 48.71 & 62.19 & 465.05 & 798.75 & 169.88 & 155.89 & 1409.76 & 1561.55 & 3502.94 & 3654.73 
\end{tblr}
\caption{Modeling results of Numerical Case Study 2, where $\mr{DOM}$ is the design optimization problem results from solving Eq.~\eqref{equ:design_optimization} and $\mr{WFP}$ are the results of solving Eq.~\eqref{equ:opt_problem_hydraulics}. }
\label{tab:overall_costs}
\end{table}

\begin{table}
\centering
\begin{tblr}{
  hline{1-2,5} = {-}{},
}
Case Study & Energy & Material  & Foundation  & $C_t - C_p$  \\
(a)        & 0.0\%      & 30.8\%       & 12.7\%         & 12.0\%      \\
(b)        & 0.0\%      & 0.0\%        & 0.0\%          & 0.0\%       \\
(c)        & -27.6\%    & 41.7\%       & -8.9\%         & 9.8\%       
\end{tblr}
\caption{Cost variation between two simulations tested, where positive values indicate that the design optimization problem provide a percentual cost reduction compared to the manual design approach.}
\label{tab:percentual_costs}
\end{table}

\section{Discussion} \label{sec:discussion}
All MAE errors between modeled results and true-based results are almost negligible, which indicates that the X-WHAT solver finds very similar, although not equal solutions to the EPANET solver. The differences are primarily due to the way the friction factor is considered. While in X-WHAT we use an explicit equation to estimate $f$ (see Eq.~\eqref{equ:friction_factor}), EPANET uses a mix within the Hagen–Poiseuille, Swamee and Jain, and a cubic interpolation formulation for different values of Reynolds number. 

All initial flow estimates for the gradient-based solver used to solve the problem were equal 10\% of the total demand. Although a non-mass-conservative guess for the solver was employed, it was still possible to find the correct answer for the flow discharges. This is an advantage of applying the method presented here in comparison with typical spreadsheet solutions of the Hardy-Cross method. For the network presented in testing case (c), it can be very tedious and time-consuming to find initial mass-conservative estimates. 


From the design optimization case studies, we found that using a leaned-on-the-ground tank, for a relatively small network (i.e., 160 inhabitants but with geometrical challenges as two nodes 10 meters higher than the tank level) was not cost-effective, see Tab.~\ref{tab:foundation_states} and Tab.~\ref{tab:pump_states}. Due to the relatively small volume and relatively high node elevations, the solver found that it was more cost-effective to build an elevated tank. This is a typical solution applied in practice in cases like that, and it was interesting that the solver found it. However, finding a solution like that was only possible when the wind forces acting on the lateral of the tank were considered and by transforming the forces acting on the tank into foundation and material costs. With the optimization procedure, costs could have been reduced 12\% for testing case (a) (see Tab.~\ref{tab:overall_costs}) simply by elevating the tank from the ground 16.07 m, as shown in Tab.~\ref{tab:foundation_states}.

The other testing cases (b) and (c) are larger networks. For both cases, it was more cost-effective to use a ground tank due to the large wind forces and the foundation cost associated with an elevated tank solution. For the testing case (b), designing the tank with trial and error and looking for a minimum tank depth to satisfy the pressure constraints resulted in the exact same solution found by the X-WHAT model. However, as shown in testing case (c), the same procedure would have resulted in an increase in costs of 151,790 USD, which represents 9.8\% of the overall cost, indicating that choosing the minimum head to satisfy the minimum pressure in the network might not be the most cost-effective solution. The non-linearities in the foundation and tank material costs make the problem of deciding the tank depths and height from the ground, especially for cases with more than 1 tank, a complex task to approach with trial and error. The optimization tool can then facilitate the design procedure.



\subsection{Limitation of this Modeling Approach}
The X-WHAT tool is a simplified Excel tool to solve steady-state WDN problems. It has the capacity to deal with relatively small networks with a maximum of 200 links. This limitation is raised by a bound in the number of maximum variables that the Excel solver supports. Although the solver packages presented in \citep{OpenSolver} find a solution to this issue, we decided not to require any other software package in lieu of Excel. In addition, having 200 links might be outside the scope of using this tool for teaching and for solving relatively small WDN systems, which is the focus of X-WHAT model.

The problem of designing the pipes, while maintaining energy and mass conservation, appears untractable to Excel basic solvers. We have extensively tried to use the simple gradient-based solvers of Excel to solve the pipeline design problem with no success. The pipeline optimization design problem has been solved in the literature with branch-and-cut algorithms that deal with mixed integer problems, increasing the complexity of such a simple tool as the one presented in this paper. However, the addition of this capability is desirable for future studies and applications of this tool.

\subsection{Advantages of the X-WHAT}
The file is less than 2 mb and works in Excel versions above 2013. We developed the tool using only Excel built-in functions to solve the WDN Water Flow Problem. Users and students are generally familiar with spreadsheet calculations and Excel functions, making it relatively easier to debug the functions and understand the steps required to solve the problems. Hence, the tool does not require coding expertise and is an advantage compared to other tools presented in the literature. All cells of input data are commented and step-by-step material on how to use the tool is provided in the supplemental information, aiding users in how to run the model. The creation of incidence matrices can be performed by flow direction algorithms, but more importantly, can be derived in class for relatively small networks; which is often the case of hydraulic exams and examples taught in classes.

The tool allows a rapid sensitivity analysis, and users can assess the effect of different demands, or lack of them, in the network, as well as assess the cost effects of different network configurations. Another interesting application of the tool is to use the hydraulic solver to calibrate the rugosity of the pipes, given a set of known pressures and/or flows in the network. Although not scoped in this paper, this analysis can be easily adapted in the tool. By changing the cost function to an error metric such as MAE or root-mean-square error within modeled and observed pressures or flows, users can define a new optimization problem with the rugosity of the pipes as the decision variable. By having the WDN numerical solver in a simple yet user-effective way, several other applications can arise.

\subsection{Tentative Class Organization}
To allow for wider application in engineering classes, we attempt to provide a tentative class organization to introduce and solve the problems presented in this paper. The methods and results presented here encompass all materials required for teaching purposes. The organization of the class is as follows


\begin{enumerate}
    \item \textit{Introduction to WDN Modeling and Design:} Two 1-h classes of fundamentals of WDN Modeling, including head loss modeling, fundamental equations of energy and mass conservation, and network topology modeling. 
    \item \textit{Solving a Looped-Network Manually:} One 1-h class for solving testing case (a), manually, step-by-step in the blackboard.
    \item \textit{Applying X-WHAT Solver to a larger network:} One 1-h class to explain to the students how to use the X-WHAT tool and apply the tool to solve testing case (a), solved manually in the previous class, and testing case (b). A homework assignment to solve testing case (c).
    \item \textit{Optimizing Tank Design:} One 1-h class to introduce design optimization and to solve test case (a) by trial and error and optimization. Homework assigned to solve testing case (c).
\end{enumerate}

%



In addition, we provide a unit design for this by applying the Understanding by Design (UbD) approach proposed in \citep{wiggins2005understanding}. UbD is a framework for designing curriculum units with a focus on student understanding and performance. This unit design can be adapted or expanded to align with the instructors' objectives, class size, and available resources. However, we are confident that it provides valuable insights into the practical application of the methods and tools proposed in our paper within an engineering classroom setting. 

\begin{landscape}
\begin{table}[]
\caption{Unit Design using UbD for an Engineering Classroom on the Design and Simulation of Looped WDNs using X-WHAT Solver}
\label{tab:Teaching_UbD}
\resizebox{\columnwidth}{!}{%
\begin{tabular}{|ll|}
\hline
\rowcolor[HTML]{BFBFBF} 
\multicolumn{2}{|c|}{\cellcolor[HTML]{BFBFBF}\textbf{STAGE 1 --- DESIRED RESULTS}}                                                                                                                         \\ \hline
\rowcolor[HTML]{FEFEFE} 
\multicolumn{2}{|l|}{\cellcolor[HTML]{FEFEFE}\begin{tabular}[c]{@{}l@{}}\textit{\textbf{Established Goals:}}\\ 
- Develop a comprehensive understanding of water distribution network (WDN) modeling and design principles.\\ 
- Master the application of fundamental equations of energy and mass conservation in WDN analysis.\\ 
- Master the design factors and structural aspects influencing the design process and optimization of WDNs.\\ 
- Acquire the skills to utilize the X-WHAT Solver for analyzing and optimizing WDNs of varying scales and complexities.\end{tabular}}                                                                                                                                   \\ \hline
\multicolumn{1}{|l|}{\begin{tabular}[c]{@{}l@{}}\textit{\textbf{Understandings:}}\\ 
\textit{Students will understand that:}\\ 
- Role of WDNs as an important infrastructure system, part of our urban development\\
- The complexity inherent in designing, operating, and management of WDNs.\\ 
- Insight into the significance of network topology in designing efficient water distribution systems.\end{tabular}}                                                                                                                                             &{\begin{tabular}[c]{@{}l@{}}\textit{\textbf{Essential Questions:}}\\ 
- What are the key equations and principles governing energy and mass conservation in WDNs?\\ - What are the design factors in WDNs?\\ 
- How to design WDNs and optimize these factors?\\ 
- How does the structural aspect affect the design process of WDN?\\ 
- How to formulate a design optimization problem that merges fundamental concepts of hydraulics \\ \;\;\; and structural design for WDNs?\\ 
- How to utilize the X-WHAT Solver for different WDN networks with different topologies and scales?\end{tabular}}                                                                                                                                                                                            \\ \hline
\rowcolor[HTML]{FEFEFE} 
\multicolumn{1}{|l|}{\cellcolor[HTML]{FEFEFE}\begin{tabular}[c]{@{}l@{}}\textit{\textbf{Knowledge:}}\\ 
\textit{Students will know:}\\ 
- Knowledge of the key equations and principles governing energy and mass conservation in WDNs.\\ - Understanding of the design factors and structural considerations influencing WDN design and optimization.\\ 
- Familiarity with the formulation of design optimization problems that merge hydraulic concepts  with structural  \\ \;\;\; design principles for WDNs.\\ 
- Simulate the hydraulics of the system using X-WHAT Solver, as well as solve the design optimization problem  \\ \;\;\; formulated.\end{tabular}}   

& \begin{tabular}[c]{@{}l@{}}\textbf{\textit{Skills:}}\\ 
\textit{Students will be able to:}\\
- Ability to apply fundamental equations and principles to analyze energy and mass conservation in WDNs.\\ 
- Skill in considering design factors and structural aspects to optimize WDN performance.\\ 
- Navigating and using the X-WHAT Solver\end{tabular}                                                                                                                           \\ \hline
\rowcolor[HTML]{BFBFBF} 
\multicolumn{2}{|c|}{\cellcolor[HTML]{BFBFBF}\textbf{STAGE 2 --- ASSESSMENT EVIDENCE}}                                                                                                                            \\ \hline
\rowcolor[HTML]{FEFEFE} 
\multicolumn{1}{|l|}{\cellcolor[HTML]{FEFEFE}\begin{tabular}[c]{@{}l@{}}\textbf{\textit{Performance Tasks:}}\\ 
- \textbf{Task 1:} Solving Test Case (a) Manually on the Board: \\ \;\;\; \textbf{Description and Criteria of Assessment:} \\ \;\;\; Students will take turns to solve the looped-network problem manually on the blackboard, demonstrating their \\ \;\;\;  understanding of fundamental equations and principles governing energy and mass conservation in WDNs.\\ 
- \textbf{Task 2:} Application and Utilization of X-WHAT Solver: \\ \;\;\; \textbf{Description and Criteria of Assessment: } \\ \;\;\; Students will utilize the X-WHAT Solver to analyze Test Case (c) as homework, demonstrating their \\ \;\;\;  ability to apply advanced tools for WDN analysis.\\ 
- \textbf{Task 3:} Design Optimization Problem: \\ \;\;\; \textbf{Description and Criteria of Assessment:} \\ \;\;\; 
Students will utilize X-WHAT Solver to integrate hydraulic concepts with structural design principles \\ \;\;\;  for Test Case (c) and solve the design optimization problem as the second homework, showcasing their ability \\ \;\;\; to merge theoretical knowledge into practical applications.\end{tabular}}       

& \begin{tabular}[c]{@{}l@{}}\textit{\textbf{Other Evidence:}}\\ 
-\textbf{ In Classroom Ongoing Discussion} while Solving the Test Cases.\\ - \textbf{During Lecture Questionnaire via Online App/Tool:}  \\ \;\;\; Students complete short questionnaires during lectures to assess their understanding of the concepts \\ \;\;\; covered in class.\\ 
\textbf{- Think-Pair-Share:} \\ \;\;\; Students respond to the results obtained from the test cases solved in class, discuss with a partner/group, \\ \;\;\;  and share with the class.\end{tabular}      \\ \hline
\rowcolor[HTML]{BFBFBF} 
\multicolumn{2}{|c|}{\cellcolor[HTML]{BFBFBF}\textbf{STAGE 3 --- LEARNING PLAN}}   \\ \hline
\rowcolor[HTML]{FEFEFE} 
\multicolumn{2}{|l|}{\cellcolor[HTML]{FEFEFE}\begin{tabular}[c]{@{}l@{}}\textbf{\textit{Learning Activities:}}\\ 
1. Pre-first class survey about students' majors, reasons to take the class, their expected outcomes, and their background knowledge. \textbf{(W, T)}\\ 
2. Lecture: Introduction to WDN Modeling - 1 hour\\ 
\;\;\; - Start the unit by highlighting the importance of WDNs and some statistics on real-world networks. \textbf{(H)} - 10 minutes\\ 
\;\;\; -  Discuss real-world engineering problems related to WDNs. \textbf{(W)} - 5 minutes\\ 
\;\;\; - Introduce the fundamentals of WDN Modeling, including head loss modeling, fundamental equations of energy and mass conservation, and network topology modeling. \textbf{(E)} - 30 minutes\\ 
\;\;\; - Reflect on the complexity associated with simulating the WDNs and how to overcome it through open discussion. \textbf{(R, E) }- 10 minutes\\ 
\;\;\; - During the Lecture Questionnaire via an online App/tool.: Students complete short questionnaires during the lecture to assess their understanding of the introduced concepts. \textbf{(E)} - 5 minutes\\ 
3. Lecture: Lopped WDN Network Simulation - 1 hour\\ 
\;\;\; - Performance Task 1: Solve Test Case (a) manually on the board by making the students take turns to solve it step-by-step. \textbf{(E)} (45 minutes)\\ 
\;\;\; - Think-Pair-Share: Divide students into groups of two or more and ask them to discuss the results together then share with the whole group. \textbf{(E, R)} (15 minutes)\\ 
4.  Lecture: Navigating and Using X-WHAT Solver - 1 hour\\ - Navigate the X-WHAT Solver in front of the students and explain what are the inputs needed and also outputs. \textbf{(E) }- 15 minutes\\ 
\;\;\; - Simulate Test Case (a) using the solver then compare the results with the results obtained from the previous lecture. \textbf{(E, R) }- 15 minutes\\ 
\;\;\; - Show students Test Case (b) and have an open discussion about the expected differences while solving this test case in comparison to Test Case (a). \textbf{(E, R)} - 10 minutes\\ 
\;\;\; - Simulate Test Case (b) using X-WHAT Solver.\\ 
5. Performance Task 2: Application and Utilization of X-WHAT Solver on Test Case (c) - Homework. \textbf{(E1)}\\ 
\;\;\; - Students are required to perform the task individually.\\ 
6. Provide feedback on performance task 2 and address common mistakes through an announcement or an email sent to students. \textbf{(R)}\\ 
7. Lecture: Design Optimization Problem and Solving it Using X-WHAT Solver - 1 hour\\ 
\;\;\; - Introduce the optimization problem: objective function, decision variables, and problem formulation taking into account the structural design aspects. \textbf{(E)} - 20 minutes\\ 
\;\;\; - Explain how to utilize the X-WHAT Solver to solve this problem through solving Test Cases (a) and (b). \textbf{(E, R)} - 30 minutes\\ 
\;\;\; - Open discussion on results \textbf{(E, R)} - 10 minutes \\ 
8. Performance Task 3: Design Optimization Problem on Test Case (c) - Homework. \textbf{(E1)}\\ 
\;\;\; - Students are required to perform the task individually.\\ 
9. Provide feedback on performance task 3 and address common mistakes through an announcement or an email to students. \textbf{(R)}\end{tabular}} \\ \hline
\end{tabular}
}
\end{table}
\end{landscape}

\section{Conclusions} \label{sec:conclusions}
In this paper, we develop an Excel tool to solve the steady-state flow distribution of an incompressible fluid in water distribution networks with nodes and links. Two problems were investigated: the WDN Water Flow Problem, and the WDN Design and Optimization Modeling. The first finds the correct flow values and directions in a given network with known geometry, tank, and pipe information, subject to fundamental conservations of mass and energy. The second, considering the same conservation constraints finds the near-optimal tank designs for a given network with known pipe information that minimizes the overall network cost. This cost considers the pumping energy to fill the tank, the tank material, and the tank foundation cost, which is calculated considering the weight of the tank and the wind forces acting on its lateral surface. 

The results of the developed hydraulic model for testing cases (a) and (b) had mean average errors of pressures and flows below 0.08 $\mr{L\cdot s^{-1}}$ or $\mr{m}$ for flows and pressures compared to EPANET, indicating acceptable results. For testing case (c), compared to the reported results of the literature, the mean average error of flows was less than 0.1\% of the maximum flow.  

The results of the optimization technique presented in this paper allowed a cost reduction of 12 and 9.8\% for testing cases (a) (i.e., a 2-loops, 1-tank network supplying drinking water to a small community of 160 inhabitants) and (b) (i.e., a 8-loops, 2-tanks, and 21 links network supplying drinking water for a population of 44,416 ). In the latter, costs had a reduction of 151,790 USD simply by optimizing the tank depths aiming to minimize the costs, to allow pressure constraints.

The numerical modeling results of optimizing the network of the testing case (a) suggest that elevated tanks are more cost-effective for WDN with relatively low demand (i.e., the tank volume was 17.28 $\mr{m^3}$, compared to the tanks of the testing case (c) of 3,572.03 and 1,224.89 $\mr{m^3}$) while facing relatively high topographic elevations. It was cost-effective to raise the tank 16 m and increase its diameter to 2.15 m, compared to a leaned-on-the-ground tank with a smaller diameter of 1.03 m. For both testing cases (b) and (c), however, due to the relatively higher tank volumes, the optimized solution was the tanks leaned on the ground. The solutions presented in this paper, interestingly, restate the common engineering practice performed in the cases of tanks to meet small and large demands.

Modeling, design, and simulation of WDN is possible using a simple spreadsheet tool such as Excel. The tool is applicable not only for teaching purposes but also for hydraulic simulation in a variety of WDN configurations. The gradient-based  GRG solver from Excel is able to provide, within a reasonable initial estimate of the decision variables, correct flow answers compared to the literature and EPANET. The design optimization problem shows that important cost reductions can be achieved if a more complete cost function is considered. By providing our open-source tool, the X-WHAT, that do not require any other software package other than Excel or any coding expertise, it is possible to advance spreadsheet-based WDN hydraulic simulations to a wider audience.  

The toolbox can be easily adapted for other problems such as calibrating network parameters for a variety of observed input such as observed pressures and flows at particular nodes and pipes. In addition, future editions of the model can include valve and pump operations and the headloss model can be enhanced by including minor losses and singularities. Although the tool was applied to drinking water systems, it is adapted to other fields such as oil and gas networks since it is allowed to change the fluid parameters in the toolbox input data. Overall, the toolbox is simple and yet can provide a variety of analysis as presented in this paper and in the supplemental material. Future advances to make the tool more accessible not only to undergraduate and graduate students, but also for the water resources engineering community are warranted.

\section{Data Availability Statement}
Some or all data, models, or code generated or used during the
study are available in a repository or online in accordance with
funder data retention policies. All software, figures, and data can be freely downloaded in \citep{ETHA_Clone}. The supplemental material also contains a user manual showcasing step-by-step how to use the developed tool.

\section*{Supplemental Materials}
Supplementary data related to this article can be found at \url{https://github.com/marcusnobrega-eng/ETHA---Clone}. 

\noindent Program size: Approximately 2 MB

\noindent Required License: Excel version 2013 or above

\noindent Availability: Open source (github license)


\bibliographystyle{cas-model2-names} 
\bibliography{Ref.bib}

\begin{thebibliography}{35}
\expandafter\ifx\csname natexlab\endcsname\relax\def\natexlab#1{#1}\fi
\providecommand{\url}[1]{\texttt{#1}}
\providecommand{\href}[2]{#2}
\providecommand{\path}[1]{#1}
\providecommand{\DOIprefix}{doi:}
\providecommand{\ArXivprefix}{arXiv:}
\providecommand{\URLprefix}{URL: }
\providecommand{\Pubmedprefix}{pmid:}
\providecommand{\doi}[1]{\href{http://dx.doi.org/#1}{\path{#1}}}
\providecommand{\Pubmed}[1]{\href{pmid:#1}{\path{#1}}}
\providecommand{\bibinfo}[2]{#2}
\ifx\xfnm\relax \def\xfnm[#1]{\unskip,\space#1}\fi
\bibitem[{Adedeji et~al.(2017)Adedeji, Hamam, Abe and
  Abu-Mahfouz}]{adedeji_spreadsheet_2017}
\bibinfo{author}{Adedeji, K.B.}, \bibinfo{author}{Hamam, Y.},
  \bibinfo{author}{Abe, B.T.}, \bibinfo{author}{Abu-Mahfouz, A.M.},
  \bibinfo{year}{2017}.
\newblock \bibinfo{title}{A spreadsheet tool for the analysis of flows in
  small-scale water piping networks}, in: \bibinfo{booktitle}{2017 {IEEE} 15th
  {International} {Conference} on {Industrial} {Informatics} ({INDIN})},
  \bibinfo{publisher}{IEEE}, \bibinfo{address}{Emden}. pp.
  \bibinfo{pages}{1213--1218}.
\newblock \URLprefix \url{http://ieeexplore.ieee.org/document/8104947/},
  \DOIprefix\doi{10.1109/INDIN.2017.8104947}.
\bibitem[{Awe et~al.(2019)Awe, Okolie and Fayomi}]{awe_review_2019}
\bibinfo{author}{Awe, O.}, \bibinfo{author}{Okolie, S.},
  \bibinfo{author}{Fayomi, O.}, \bibinfo{year}{2019}.
\newblock \bibinfo{title}{Review of {Water} {Distribution} {Systems}
  {Modelling} and {Performance} {Analysis} {Softwares}}, in:
  \bibinfo{booktitle}{Journal of {Physics}: {Conference} {Series}}, p.
  \bibinfo{pages}{022067}.
\newblock \URLprefix
  \url{https://iopscience.iop.org/article/10.1088/1742-6596/1378/2/022067},
  \DOIprefix\doi{10.1088/1742-6596/1378/2/022067}.
\bibitem[{Barrón~Corvera et~al.(2021)Barrón~Corvera, Alvarado~Medellín,
  Bautista~Capetillo and Badillo~Almaraz}]{barron_corvera_herramienta_2021}
\bibinfo{author}{Barrón~Corvera, A.}, \bibinfo{author}{Alvarado~Medellín,
  P.}, \bibinfo{author}{Bautista~Capetillo, C.F.},
  \bibinfo{author}{Badillo~Almaraz, H.}, \bibinfo{year}{2021}.
\newblock \bibinfo{title}{Herramienta informática para diseño de redes
  hidráulicas presurizadas}.
\newblock \bibinfo{journal}{Acta Universitaria} \bibinfo{volume}{31},
  \bibinfo{pages}{1--17}.
\newblock \URLprefix
  \url{https://www.actauniversitaria.ugto.mx/index.php/acta/article/view/3093},
  \DOIprefix\doi{10.15174/au.2021.3093}.
\bibitem[{Bragalli et~al.(2012)Bragalli, D’Ambrosio, Lee, Lodi and
  Toth}]{bragalli2012optimal}
\bibinfo{author}{Bragalli, C.}, \bibinfo{author}{D’Ambrosio, C.},
  \bibinfo{author}{Lee, J.}, \bibinfo{author}{Lodi, A.}, \bibinfo{author}{Toth,
  P.}, \bibinfo{year}{2012}.
\newblock \bibinfo{title}{On the optimal design of water distribution networks:
  a practical minlp approach}.
\newblock \bibinfo{journal}{Optimization and Engineering} \bibinfo{volume}{13},
  \bibinfo{pages}{219--246}.
\bibitem[{Brkić(2017)}]{brkic_spreadsheet-based_2017}
\bibinfo{author}{Brkić, D.}, \bibinfo{year}{2017}.
\newblock \bibinfo{title}{Spreadsheet-{Based} {Pipe} {Networks} {Analysis} for
  {Teaching} and {Learning} {Purpose}}.
\newblock \bibinfo{type}{preprint}. Open Science Framework.
\newblock \URLprefix \url{https://osf.io/7ynvw},
  \DOIprefix\doi{10.31219/osf.io/7ynvw}.
\bibitem[{Cassiolato et~al.(2023)Cassiolato, de~Carvalho and
  Ravagnani}]{cassiolato2023minlp}
\bibinfo{author}{Cassiolato, G.H.B.}, \bibinfo{author}{de~Carvalho, E.P.},
  \bibinfo{author}{Ravagnani, M.A.d.S.S.}, \bibinfo{year}{2023}.
\newblock \bibinfo{title}{An minlp model for the minimization of installation
  and operational costs in water distribution networks}.
\newblock \bibinfo{journal}{Acta Scientiarum. Technology} \bibinfo{volume}{45},
  \bibinfo{pages}{e59993--e59993}.
\bibitem[{Creaco et~al.(2019)Creaco, Campisano, Fontana, Marini, Page and
  Walski}]{creaco2019real}
\bibinfo{author}{Creaco, E.}, \bibinfo{author}{Campisano, A.},
  \bibinfo{author}{Fontana, N.}, \bibinfo{author}{Marini, G.},
  \bibinfo{author}{Page, P.}, \bibinfo{author}{Walski, T.},
  \bibinfo{year}{2019}.
\newblock \bibinfo{title}{Real time control of water distribution networks: A
  state-of-the-art review}.
\newblock \bibinfo{journal}{Water research} \bibinfo{volume}{161},
  \bibinfo{pages}{517--530}.
\bibitem[{Cross(1936)}]{cross1936analysis}
\bibinfo{author}{Cross, H.}, \bibinfo{year}{1936}.
\newblock \bibinfo{title}{Analysis of flow in networks of conduits or
  conductors}.
\newblock \bibinfo{journal}{University of Illinois. Engineering Experiment
  Station. Bulletin; no. 286} .
\bibitem[{Demir et~al.(2018)Demir, Manav~Demir and Karadeniz}]{demir_ms_2018}
\bibinfo{author}{Demir, S.}, \bibinfo{author}{Manav~Demir, N.},
  \bibinfo{author}{Karadeniz, A.}, \bibinfo{year}{2018}.
\newblock \bibinfo{title}{An {MS} {Excel} tool for water distribution network
  design in environmental engineering education}.
\newblock \bibinfo{journal}{Computer Applications in Engineering Education}
  \bibinfo{volume}{26}, \bibinfo{pages}{203--214}.
\newblock \URLprefix
  \url{https://onlinelibrary.wiley.com/doi/10.1002/cae.21870},
  \DOIprefix\doi{10.1002/cae.21870}.
\bibitem[{Eliades et~al.(2016)Eliades, Kyriakou, Vrachimis and
  Polycarpou}]{eliades_EPANET-matlab_2016}
\bibinfo{author}{Eliades, D.G.}, \bibinfo{author}{Kyriakou, M.},
  \bibinfo{author}{Vrachimis, S.}, \bibinfo{author}{Polycarpou, M.M.},
  \bibinfo{year}{2016}.
\newblock \bibinfo{title}{{EPANET}-{MATLAB} {Toolkit}: {An} {Open}-{Source}
  {Software} for {Interfacing} {EPANET} with {MATLAB}}, in:
  \bibinfo{booktitle}{Proc. 14th {International} {Conference} on {Computing}
  and {Control} for the {Water} {Industry} ({CCWI})}, \bibinfo{address}{The
  Netherlands}.
\newblock \DOIprefix\doi{10.5281/zenodo.831493}.
\bibitem[{Giustolisi et~al.(2011)Giustolisi, Savić, Berardi and
  Laucelli}]{giustolisi_excel-based_2011}
\bibinfo{author}{Giustolisi, O.}, \bibinfo{author}{Savić, D.A.},
  \bibinfo{author}{Berardi, L.}, \bibinfo{author}{Laucelli, D.},
  \bibinfo{year}{2011}.
\newblock \bibinfo{title}{{AN} {EXCEL}-{BASED} {SOLUTION} {TO} {BRING} {WATER}
  {DISTRIBUTION} {NETWORK} {ANALYSIS} {CLOSER} {TO} {USERS}} .
\bibitem[{Gokyay(2020)}]{gokyay_easy_2020}
\bibinfo{author}{Gokyay, O.}, \bibinfo{year}{2020}.
\newblock \bibinfo{title}{An easy {MS} {Excel} software to use for water
  distribution system design: {A} real case distribution network design
  solution}.
\newblock \bibinfo{journal}{Journal of Applied Water Engineering and Research}
  \bibinfo{volume}{8}, \bibinfo{pages}{290--297}.
\newblock \URLprefix
  \url{https://www.tandfonline.com/doi/full/10.1080/23249676.2020.1831975},
  \DOIprefix\doi{10.1080/23249676.2020.1831975}.
\bibitem[{{Gomes Jr.}(2024)}]{ETHA_Clone}
\bibinfo{author}{{Gomes Jr.}}, \bibinfo{year}{2024}.
\newblock \bibinfo{title}{Etha-clone v.0.0.1}.
\newblock
  \bibinfo{howpublished}{\url{https://github.com/marcusnobrega-eng/ETHA---Clone}}.
\bibitem[{Huddleston et~al.(2004a)Huddleston, Alarcon and
  Chen}]{huddleston_water_2004}
\bibinfo{author}{Huddleston, D.H.}, \bibinfo{author}{Alarcon, V.J.},
  \bibinfo{author}{Chen, W.}, \bibinfo{year}{2004}a.
\newblock \bibinfo{title}{Water {Distribution} {Network} {Analysis} {Using}
  {Excel}}.
\newblock \bibinfo{journal}{Journal of Hydraulic Engineering}
  \bibinfo{volume}{130}, \bibinfo{pages}{1033--1035}.
\newblock \URLprefix
  \url{https://ascelibrary.org/doi/10.1061/%28ASCE%290733-9429%282004%29130%3A10%281033%29},
  \DOIprefix\doi{10.1061/(ASCE)0733-9429(2004)130:10(1033)}.
\bibitem[{Huddleston et~al.(2004b)Huddleston, Alarcon and
  Chen}]{huddleston2004water}
\bibinfo{author}{Huddleston, D.H.}, \bibinfo{author}{Alarcon, V.J.},
  \bibinfo{author}{Chen, W.}, \bibinfo{year}{2004}b.
\newblock \bibinfo{title}{Water distribution network analysis using excel}.
\newblock \bibinfo{journal}{Journal of hydraulic engineering}
  \bibinfo{volume}{130}, \bibinfo{pages}{1033--1035}.
\bibitem[{Jewell(2001)}]{jewell_teaching_2001}
\bibinfo{author}{Jewell, T.K.}, \bibinfo{year}{2001}.
\newblock \bibinfo{title}{Teaching {Hydraulic} {Design} {Using} {Equation}
  {Solvers}}.
\newblock \bibinfo{journal}{Journal of Hydraulic Engineering}
  \bibinfo{volume}{127}, \bibinfo{pages}{1013--1021}.
\newblock \URLprefix
  \url{https://ascelibrary.org/doi/10.1061/%28ASCE%290733-9429%282001%29127%3A12%281013%29},
  \DOIprefix\doi{10.1061/(ASCE)0733-9429(2001)127:12(1013)}.
\bibitem[{von Krogh and Spaeth(2007)}]{von_krogh_open_2007}
\bibinfo{author}{von Krogh, G.}, \bibinfo{author}{Spaeth, S.},
  \bibinfo{year}{2007}.
\newblock \bibinfo{title}{The open source software phenomenon:
  {Characteristics} that promote research}.
\newblock \bibinfo{journal}{The Journal of Strategic Information Systems}
  \bibinfo{volume}{16}, \bibinfo{pages}{236--253}.
\newblock \URLprefix
  \url{https://www.sciencedirect.com/science/article/pii/S096386870700025X},
  \DOIprefix\doi{10.1016/j.jsis.2007.06.001}.
\bibitem[{Ku et~al.(2011)Ku, Fulcher and Xiang}]{ku_using_2011}
\bibinfo{author}{Ku, H.}, \bibinfo{author}{Fulcher, R.},
  \bibinfo{author}{Xiang, W.}, \bibinfo{year}{2011}.
\newblock \bibinfo{title}{Using computer software packages to enhance the
  teaching of engineering management science: {Part} 1—{Critical} path
  networks}.
\newblock \bibinfo{journal}{Computer Applications in Engineering Education}
  \bibinfo{volume}{19}, \bibinfo{pages}{26--39}.
\newblock \URLprefix
  \url{https://onlinelibrary.wiley.com/doi/abs/10.1002/cae.20286},
  \DOIprefix\doi{10.1002/cae.20286}. \bibinfo{note}{\_eprint:
  https://onlinelibrary.wiley.com/doi/pdf/10.1002/cae.20286}.
\bibitem[{Mason(2012)}]{OpenSolver}
\bibinfo{author}{Mason, A.}, \bibinfo{year}{2012}.
\newblock \bibinfo{title}{Opensolver – an open source add-in to solve linear
  and integer progammes in excel}, in: \bibinfo{editor}{Klatte, D.},
  \bibinfo{editor}{Lathi, H.J.}, \bibinfo{editor}{Schmedders, K.} (Eds.),
  \bibinfo{booktitle}{Operations Research Proceedings 2011}.
  \bibinfo{publisher}{Springer Berlin Heidelberg}. Operations Research
  Proceedings, pp. \bibinfo{pages}{401--406}.
\newblock \URLprefix \url{http://dx.doi.org/10.1007/978-3-642-29210-1_64},
  \DOIprefix\doi{10.1007/978-3-642-29210-1_64}.
  \bibinfo{note}{http://opensolver.org}.
\bibitem[{Mays(2000)}]{mays_water_2000}
\bibinfo{author}{Mays, L.W.}, \bibinfo{year}{2000}.
\newblock \bibinfo{title}{Water {Distribution} {System} {Handbook}}.
\newblock \bibinfo{edition}{1st edition} ed., \bibinfo{publisher}{McGraw-Hill
  Education}.
\newblock \URLprefix
  \url{https://www.accessengineeringlibrary.com/content/book/9780071342131}.
\bibitem[{Niazkar and Afzali(2017a)}]{niazkar_analysis_2017}
\bibinfo{author}{Niazkar, M.}, \bibinfo{author}{Afzali, S.H.},
  \bibinfo{year}{2017}a.
\newblock \bibinfo{title}{Analysis of water distribution networks using
  {MATLAB} and {Excel} spreadsheet: h-based methods: {IMPLEMENTATION} {OF}
  h-{BASED} {METHODS} {IN} {MATLAB} {AND} {EXCEL} {SPREADSHEET}}.
\newblock \bibinfo{journal}{Computer Applications in Engineering Education}
  \bibinfo{volume}{25}, \bibinfo{pages}{129--141}.
\newblock \URLprefix
  \url{https://onlinelibrary.wiley.com/doi/10.1002/cae.21786},
  \DOIprefix\doi{10.1002/cae.21786}.
\bibitem[{Niazkar and Afzali(2017b)}]{niazkar_analysis_2017-1}
\bibinfo{author}{Niazkar, M.}, \bibinfo{author}{Afzali, S.H.},
  \bibinfo{year}{2017}b.
\newblock \bibinfo{title}{Analysis of water distribution networks using
  {MATLAB} and {Excel} spreadsheet: {Q}-based methods: {IMPLEMENTATION} {OF}
  {Q}-{BASED} {METHODS} {IN} {MATLAB} {AND} {EXCEL} {SPREADSHEET}}.
\newblock \bibinfo{journal}{Computer Applications in Engineering Education}
  \bibinfo{volume}{25}, \bibinfo{pages}{277--289}.
\newblock \URLprefix
  \url{https://onlinelibrary.wiley.com/doi/10.1002/cae.21796},
  \DOIprefix\doi{10.1002/cae.21796}.
\bibitem[{Oke et~al.(2015)Oke, Ismail, Lukman, Adie, Adeosun, Umaru and
  Nwude}]{oke_statistical_2015}
\bibinfo{author}{Oke, I.A.}, \bibinfo{author}{Ismail, A.},
  \bibinfo{author}{Lukman, S.}, \bibinfo{author}{Adie, D.B.},
  \bibinfo{author}{Adeosun, O.O.}, \bibinfo{author}{Umaru, A.B.},
  \bibinfo{author}{Nwude, M.O.}, \bibinfo{year}{2015}.
\newblock \bibinfo{title}{Statistical evaluation of mathematical methods in
  solving linear theory problems: {Design} of water distribution systems}.
\newblock \bibinfo{journal}{Ife Journal of Science} \bibinfo{volume}{17},
  \bibinfo{pages}{255--267}.
\newblock \URLprefix
  \url{https://www.ajol.info/index.php/ijs/article/view/131756},
  \DOIprefix\doi{10.4314/ijs.v17i2}. \bibinfo{note}{number: 2}.
\bibitem[{Parker(2010)}]{parker_selecting_2010}
\bibinfo{author}{Parker, K.R.}, \bibinfo{year}{2010}.
\newblock \bibinfo{title}{Selecting software tools for {IS}/{IT} curricula}.
\newblock \bibinfo{journal}{Education and Information Technologies}
  \bibinfo{volume}{15}, \bibinfo{pages}{255--275}.
\newblock \URLprefix \url{https://doi.org/10.1007/s10639-010-9126-8},
  \DOIprefix\doi{10.1007/s10639-010-9126-8}.
\bibitem[{PORTO(2006)}]{porto2004hidraulica}
\bibinfo{author}{PORTO, R.d.M.}, \bibinfo{year}{2006}.
\newblock \bibinfo{title}{Hidr{\'a}ulica b{\'a}sica. 4{\textordfeminine}
  edi{\c{c}}{\~a}o}.
\newblock \bibinfo{journal}{S{\~a}o Carlos: EESC-USP, Projeto REENGE} .
\bibitem[{Rivas et~al.(2006)Rivas, Gómez-Acebo and
  Ramos}]{rivas_application_2006}
\bibinfo{author}{Rivas, A.}, \bibinfo{author}{Gómez-Acebo, T.},
  \bibinfo{author}{Ramos, J.C.}, \bibinfo{year}{2006}.
\newblock \bibinfo{title}{The application of spreadsheets to the analysis and
  optimization of systems and processes in the teaching of hydraulic and
  thermal engineering}.
\newblock \bibinfo{journal}{Computer Applications in Engineering Education}
  \bibinfo{volume}{14}, \bibinfo{pages}{256--268}.
\newblock \URLprefix
  \url{https://onlinelibrary.wiley.com/doi/10.1002/cae.20085},
  \DOIprefix\doi{10.1002/cae.20085}.
\bibitem[{Rossman(2000)}]{rossman_epanet_2000}
\bibinfo{author}{Rossman, L.A.}, \bibinfo{year}{2000}.
\newblock \bibinfo{title}{{EPANET} 2: users manual}.
\newblock \bibinfo{publisher}{Cincinnati, OH : U.S. Environmental Protection
  Agency, Risk Reduction Engineering Laboratory}.
\bibitem[{Smith and Lasdon(1992)}]{smith1992solving}
\bibinfo{author}{Smith, S.}, \bibinfo{author}{Lasdon, L.},
  \bibinfo{year}{1992}.
\newblock \bibinfo{title}{Solving large sparse nonlinear programs using grg}.
\newblock \bibinfo{journal}{ORSA Journal on Computing} \bibinfo{volume}{4},
  \bibinfo{pages}{2--15}.
\bibitem[{Sonaje and Joshi(2015)}]{sonaje_review_2015}
\bibinfo{author}{Sonaje, N.P.}, \bibinfo{author}{Joshi, M.G.},
  \bibinfo{year}{2015}.
\newblock \bibinfo{title}{A {REVIEW} {OF} {MODELING} {AND} {APPLICATION} {OF}
  {WATER} {DISTRIBUTION} {NETWORKS} ({WDN}) {SOFTWARES}}.
\bibitem[{Thakur(2012)}]{thakur_limited_2012}
\bibinfo{author}{Thakur, D.}, \bibinfo{year}{2012}.
\newblock \bibinfo{title}{A limited revolution — {The} distributional
  consequences of {Open} {Source} {Software} in {North} {America}}.
\newblock \bibinfo{journal}{Technological Forecasting and Social Change}
  \bibinfo{volume}{79}, \bibinfo{pages}{244--251}.
\newblock \URLprefix
  \url{https://www.sciencedirect.com/science/article/pii/S0040162511002174},
  \DOIprefix\doi{10.1016/j.techfore.2011.10.003}.
\bibitem[{Türkkan et~al.(2020)Türkkan, Eryılmaz~Türkkan and
  Yılmaz}]{turkkan_visual_2020}
\bibinfo{author}{Türkkan, Y.A.}, \bibinfo{author}{Eryılmaz~Türkkan, G.},
  \bibinfo{author}{Yılmaz, H.}, \bibinfo{year}{2020}.
\newblock \bibinfo{title}{A visual application for teaching pipe flow
  optimization in engineering curricula}.
\newblock \bibinfo{journal}{Computer Applications in Engineering Education}
  \bibinfo{volume}{28}, \bibinfo{pages}{154--159}.
\newblock \URLprefix
  \url{https://onlinelibrary.wiley.com/doi/10.1002/cae.22181},
  \DOIprefix\doi{10.1002/cae.22181}.
\bibitem[{Wahba(2015)}]{wahba_improved_2015}
\bibinfo{author}{Wahba, E.M.}, \bibinfo{year}{2015}.
\newblock \bibinfo{title}{An improved computational algorithm for teaching
  hydraulics of branching pipes in engineering curricula: {Hydraulics} of
  {Branching} {Pipes}}.
\newblock \bibinfo{journal}{Computer Applications in Engineering Education}
  \bibinfo{volume}{23}, \bibinfo{pages}{537--541}.
\newblock \URLprefix
  \url{https://onlinelibrary.wiley.com/doi/10.1002/cae.21624},
  \DOIprefix\doi{10.1002/cae.21624}.
\bibitem[{Walski et~al.(2003)Walski, Chase, Savic, Grayman, Beckwith and
  Koelle}]{walski_advanced_2003}
\bibinfo{author}{Walski, T.M.}, \bibinfo{author}{Chase, D.V.},
  \bibinfo{author}{Savic, D.A.}, \bibinfo{author}{Grayman, W.},
  \bibinfo{author}{Beckwith, S.}, \bibinfo{author}{Koelle, E.},
  \bibinfo{year}{2003}.
\newblock \bibinfo{title}{Advanced {Water} {Distribution} {Modeling} and
  {Management}}.
\newblock \bibinfo{publisher}{Civil and Environmental Engineering and
  Engineering Mechanics Faculty Publications. Paper 18.},
  \bibinfo{address}{Waterbury, CT}.
\bibitem[{White(1966)}]{whiteFluidMechanics1966a}
\bibinfo{author}{White, F.M.}, \bibinfo{year}{1966}.
\newblock \bibinfo{title}{Fluid Mechanics}.
\bibitem[{Wiggins and McTighe(2005)}]{wiggins2005understanding}
\bibinfo{author}{Wiggins, G.}, \bibinfo{author}{McTighe, J.},
  \bibinfo{year}{2005}.
\newblock \bibinfo{title}{Understanding by design}.
\newblock \bibinfo{publisher}{Ascd}.

\end{thebibliography}





\end{document}